\documentclass[prb,tighten]{revtex4}

\usepackage{amssymb,amsmath}

\usepackage{graphicx}% Include figure files

\begin{document}
\title{Transport coefficients for inelastic Maxwell mixtures}
\author{Vicente Garz\'{o}}
 \email[E-mail: ]{vicenteg@unex.es}
\address{Departamento de F\'{\i}sica, Universidad de Extremadura, E-06071
Badajoz, Spain}
\author{Antonio Astillero} \email[E-mail: ]{aavivas@unex.es}
\address{Departamento de Inform\'atica, Centro Universitario de M\'erida, Universidad de Extremadura, E-06800 M\'erida, Spain}

%\tighten

\begin{abstract}

The Boltzmann equation for inelastic Maxwell models is used to determine the Navier-Stokes transport coefficients of a granular binary mixture in $d$ dimensions. The Chapman-Enskog method is applied to solve the Boltzmann equation for states near the (local) homogeneous cooling state. The mass, heat, and momentum fluxes are obtained to first order in the spatial gradients of the hydrodynamic fields, and the corresponding transport coefficients are identified. There are seven relevant transport coefficients: the mutual diffusion, the pressure diffusion, the thermal diffusion, the shear viscosity, the Dufour coefficient, the pressure energy coefficient, and the thermal conductivity. All these coefficients are {\em exactly} obtained in terms of the coefficients of restitution and the ratios of mass, concentration, and particle sizes. The results are compared with known transport coefficients of inelastic hard spheres obtained analytically in the leading Sonine approximation and by means of Monte Carlo simulations. The comparison shows a reasonably good agreement between both interaction models for not too strong dissipation, especially in the case of the transport coefficients associated with the mass flux.


Running title:  Inelastic Maxwell mixtures

\end{abstract}

%\draft
\date{\today}
\maketitle
\section{Introduction}
\label{sec1}

The evaluation of the transport coefficients from the Boltzmann equation for inelastic hard spheres (IHS) is quite involved. In fact, to get explicit results one usually considers the leading order in a Sonine polynomial expansion of the velocity distribution function. These difficulties increase when one considers multicomponent systems since not  only the number of transport coefficients is larger than for a single gas but  they are also functions of more parameters such as composition, masses, sizes, and the coefficients of restitution. As in the elastic case, a possible way to partially overcome these problems is to consider other interaction models that simplify the complex mathematical structure of the Boltzmann collision integrals for IHS. For this reason, the so-called inelastic Maxwell models (IMM) have been widely used in the past few years as a toy  model to characterize the influence of the inelasticity of  collisions on the physical properties of granular fluids. The IMM share with elastic Maxwell molecules the property   that the collision rate is velocity independent but, on the other hand, their collision rules are the same as for IHS. In this sense, although these IMM do not correspond to any microscopic potential interaction, it has been shown by several authors \cite{BCG00,CCG00,NK00,EB02,BMP02,KN02,BC03,BCT03,BE04} that the cost of sacrificing physical realism is in part compensated by the amount of exact analytical results.

Most of the studies carried out by considering IMM have been devoted to homogeneous states, especially in the analysis of the overpopulated high energy tails. \cite{EB02,KN02,EB02bis,BC03,BCT03,SE03} However, much less is known in the case of {\em inhomogeneous} situations and, more specifically, on the dependence of the transport coefficients on dissipation. For a monocomponent granular gas subjected to simple shear flow, the IMM have been used to calculate the rheological properties (shear and normal stresses) in three dimensions. \cite{C01} More recently, this study has been extended \cite{G03} to multicomponent systems, the exact results of IMM showing a close agreement with those obtained analytically for IHS in the first Sonine approximation \cite{ MG02} and by means of Monte Carlo simulations. \cite{MG02,MG03} All these results are restricted to steady shear flow problems without any limitation on the strength of the shear rate. For general inhomogeneous problems and in the case of a monocomponent gas, the Boltzmann equation for IMM has been solved \cite{S03} from the Chapman-Enskog method \cite{CC70} for states near the (local) homogeneous cooling state. Explicit expressions of the Navier-Stokes transport coefficients of IMM in $d$ dimensions have been obtained for unforced systems as well as for systems driven by thermostats. In contrast to the findings of Ref.\ \onlinecite{G03}, the comparison with the transport coefficients of IHS \cite{BDKS98,GM02} shows that their dependence on inelasticity is captured by the IMM only in a mild qualitative way. This fact stimulates the determination of the exact expressions of the transport coefficients for inelastic granular mixtures.

The goal of this paper is to derive the hydrodynamic equations for a $d$-dimensional binary mixture of inelastic Maxwell gases at low-density. As in the single gas case, \cite{S03} a {\em normal} solution to the coupled set of Boltzmann equations for the two species is obtained by using the standard Chapman-Enskog method \cite{CC70} conveniently generalized to inelastic collisions. In the first order of the spatial gradients of the hydrodynamic fields, we get the corresponding Navier-Stokes hydrodynamic equations with explicit expressions for the relevant transport coefficients of the mixture. For molecular fluid mixtures, the specific set of gradients contributing to each flux is restricted by fluid symmetry, time reversal invariance (Onsager relations), and the form of entropy production. \cite{GM84} However, in the case of inelastic collisions, only fluid symmetry holds and so there is more flexibility in representing the fluxes and identifying the corresponding transport coefficients. In the case of the pressure tensor, fluid symmetry  considerations  implies that its form to first order in the gradients is the same as for the monocomponent gas. In the case of mass and heat fluxes, several different (but equivalent) choices of hydrodynamic fields can be used and some care must be taken when comparing transport coefficients coming from different representations. As in the case of IHS,\cite{GD02} we take the gradients of the mole fraction,
the (hydrostatic) pressure, the temperature, and the flow velocity as
the relevant ones. As a consequence, in this representation there are seven independent scalar transport coefficients: the mutual diffusion, the pressure diffusion and the thermal diffusion associated with the mass flux, the shear viscosity corresponding to the pressure tensor and the Dufour coefficient, the thermal conductivity, and the pressure energy coefficient associated with the heat flux. All these coefficients are given in terms of the coefficients of restitution as well as on the ratios of concentration, masses and particle sizes. In addition, as in the previous study for IHS, \cite{GD02} our theory takes into account the effect of temperature differences  (failure of energy equipartition) on the transport coefficients, leading to additional dependencies of them on the concentration.

The plan of the paper is as follows. In Section \ref{sec2} the Boltzmann equation for IMM and the macroscopic conservation laws are introduced. The model includes average collision frequencies $\omega_{rs}$ which can be freely fitted to get good agreement with IHS. Here, we fix $\omega_{rs}$ by the criterion that the cooling rates $\zeta_{rs}$ of IMM be the same as those obtained for IHS in the local equilibrium approximation. The homogenous solution of the Boltzmann equation is analyzed in Section \ref{sec3}, where the temperature ratio and the fourth cumulant (kurtosis) of the velocity distribution functions are exactly obtained. Comparison with the results obtained for IHS \cite{GD99,MG02bis} shows an excellent agreement for the temperature ratio but significant discrepancies with the fourth cumulant of IHS. Section \ref{sec4} deals with the application of the Chapman-Enskog method to get the transport coefficients of IMM. In Section \ref{sec5}, the dependence of some of these coefficients on the parameters of the system is illustrated and compared with known results derived for IHS. \cite{GD99,DG01,MG03bis,GM04} The comparison shows in general a qualitative good agreement, especially for the transport coefficients defining the mass flux. The paper ends in Section \ref{sec6} with a brief discussion on the results reported in this paper.

\section{Inelastic Maxwell models for a granular binary mixture}
\label{sec2}

Let us consider a binary mixture of inelastic Maxwell gases at low density.  The Boltzmann equation for IMM \cite{NK00,EB02,EB02bis,NK02} can be obtained from the Boltzmann equation for IHS by replacing the  rate for collisions between particles of species $r$ and $s$ by an average  velocity-independent
collision rate, which is proportional to the square root of the ``granular''  temperature $T$.  This means that a random pair of colliding particles undergoes inelastic collisions with a random impact direction. With this simplification, the velocity distribution functions $f_r({\bf r}, {\bf v};t)$ $( r=1, 2)$ of each species  satisfy the following set of nonlinear Boltzmann kinetic equations:
\begin{equation}
\label{2.1}
\left(\partial_t+{\bf v}\cdot \nabla \right)f_{r}
({\bf r},{\bf v};t)
=\sum_{s}J_{rs}\left[{\bf v}|f_{r}(t),f_{s}(t)\right] \;,
\end{equation}
where  the Boltzmann collision operator $J_{rs}\left[{\bf v}|f_{r},f_{s}\right]$ is
\begin{eqnarray}
J_{rs}\left[{\bf v}_{1}|f_{r},f_{s}\right] &=&\frac{\omega_{rs}({\bf r},t;\alpha_{rs})}{n_s({\bf r},t)\Omega_d}
\int d{\bf v}_{2}\int d\widehat{\boldsymbol{\sigma }}\nonumber\\
& & \times
\left[ \alpha_{rs}^{-1}f_{r}({\bf r},{\bf v}_{1}',t)f_{s}(
{\bf r},{\bf v}_{2}',t)-f_{r}({\bf r},{\bf v}_{1},t)f_{s}({\bf r},{\bf v}_{2},t)\right]
\;.
\label{2.2}
\end{eqnarray}
Here $n_r$ is the number density of species $r$, $\omega_{rs}\neq \omega_{sr}$ is an effective collision
frequency (to be chosen later) for collisions  of type $r$-$s$,  $\Omega_d=2\pi^{d/2}/\Gamma(d/2)$ is the total solid angle in $d$ dimensions, and $\alpha_{rs}=\alpha_{sr}\leq 1$ refers to the constant coefficient of restitution for collisions between particles of species $r$ with $s$.   In
addition, the primes on the velocities denote the initial values $\{{\bf v}_{1}^{\prime},
{\bf v}_{2}^{\prime}\}$ that lead to $\{{\bf v}_{1},{\bf v}_{2}\}$
following a binary collision:
\begin{equation}
\label{2.3}
{\bf v}_{1}^{\prime }={\bf v}_{1}-\mu_{sr}\left( 1+\alpha_{rs}
^{-1}\right)(\widehat{\boldsymbol{\sigma}}\cdot {\bf g}_{12})\widehat{\boldsymbol
{\sigma}},
\quad {\bf v}_{2}^{\prime}={\bf v}_{2}+\mu_{rs}\left(
1+\alpha_{rs}^{-1}\right) (\widehat{\boldsymbol{\sigma}}\cdot {\bf
g}_{12})\widehat{\boldsymbol{\sigma}}\;,
\end{equation}
where ${\bf g}_{12}={\bf v}_1-{\bf v}_2$ is the relative velocity of the colliding pair,
$\widehat{\boldsymbol{\sigma}}$ is a unit vector directed along the centers of the two colliding
spheres, and $\mu_{rs}=m_r/(m_r+m_s)$. The collision frequencies $\omega_{rs}$ can be seen as free parameters in the model. Its dependence on the restitution coefficients $\alpha_{rs}$ can be chosen to
optimize the agreement with  the results obtained from the Boltzmann equation  for IHS. Of course,
the choice is not unique and may depend on the property of interest.

There is another more refined version of the inelastic Maxwell model \cite{BCG00,CCG00,BC02} where the collision rate  has the same dependence on the scalar product ($ \widehat{\boldsymbol{\sigma }}\cdot \widehat{{\bf g}}_{12}$) as in the case of hard spheres. However, both versions of IMM lead to  similar results in problems as delicate as the high energy tails.\cite{EB02}   Therefore, for the sake of simplicity, here we will consider the version given by Eqs.\ (\ref{2.2}) and (\ref{2.3}).

The relevant hydrodynamic fields in a binary mixture are the number densities $n_r$, the flow velocity  ${\bf u}$, and the granular temperature $T$. They are defined in terms of the distribution $f_r$ as
\begin{equation}
\label{2.2.1}
n_r=\int d{\bf v} f_r({\bf v}),
\end{equation}
\begin{equation}
\label{2.4}
 \rho{\bf u}=\sum_r\rho_r{\bf u}_r=\sum_r\int d{\bf v}m_r{\bf v}f_r({\bf v}),
\end{equation}
\begin{equation}
\label{2.5}
nT=p=\sum_rn_rT_r=\sum_r\frac{m_r}{d}\int d{\bf v}V^2f_r({\bf v}),
\end{equation}
where $\rho_r=m_rn_r$ is the mass density of species $r$, $n=n_1+n_2$ is the total number density, $\rho=\rho_1+\rho_2$ is the
total mass density, ${\bf V}={\bf v}-{\bf u}$ is the peculiar velocity, and $p$ is the hydrostatic pressure. Furthermore, the third equality of Eq.\ (\ref{2.5}) defines the kinetic temperatures $T_r$ of each species, which measure  their mean kinetic energies.

The collision operators conserve the particle number of each species and the
total momentum, but the total energy is not conserved:
\begin{equation}
\label{2.6.1}
\int\, d{\bf v} J_{rs}[{\bf v}|f_r,f_s]=0,
\end{equation}
\begin{equation}
\label{2.6.2}
\sum_{r,s}\int d{\bf v}m_{r}{\bf v} J_{rs}[{\bf v}|f_{r},f_{s}]=0,
\end{equation}
\begin{equation}
\label{2.6}
\sum_{r,s}\int d{\bf v}\case{1}{2}m_{r}V^{2}J_{rs}
[{\bf v}|f_{r},f_{s}]=-\case{d}{2}nT\zeta \;,
\end{equation}
where $\zeta$ is identified as the ``cooling rate'' due to inelastic
collisions among all species. At a kinetic level, it is also convenient to introduce
the ``cooling rates'' $\zeta_r$ for the partial temperatures $T_r$. They are defined as
\begin{equation}
\label{2.7}
\zeta_r=\sum_s \zeta_{rs}=-\sum_s \frac{1}{dn_rT_r}\int d{\bf v}m_rV^{2}J_{rs}[{\bf v}|f_{r},f_{s}],
\end{equation}
where the second equality defines the quantities $\zeta_{rs}$. The total cooling rate $\zeta$ can be written in terms of the partial cooling rates $\zeta_r$ as
\begin{equation}
\label{2.8}
\zeta=T^{-1}\sum_rx_rT_r\zeta_r,
\end{equation}
where $x_r=n_r/n$ is the mole fraction of species $r$.

From Eqs.\ (\ref{2.2.1}) to (\ref{2.6}), the macroscopic balance equations for the binary mixture can be obtained. They are given by
\begin{equation}
D_{t}n_{r}+n_{r}\nabla \cdot {\bf u}+\frac{\nabla \cdot {\bf j}_{r}}{m_{r}}
=0\;,  \label{2.9}
\end{equation}
\begin{equation}
D_{t}{\bf u}+\rho ^{-1}\nabla \cdot {\sf P}=0\;,  \label{2.10}
\end{equation}
\begin{equation}
D_{t}T-\frac{T}{n}\sum_{r}\frac{\nabla \cdot {\bf j}_{r}}{m_{r}}+\frac{2}{dn}
\left( \nabla \cdot {\bf q}+{\sf P}:\nabla {\bf u}\right)
=-\zeta T\;. \label{2.11}
\end{equation}
In the above equations, $D_{t}=\partial _{t}+{\bf u}\cdot \nabla $ is the
material derivative,
\begin{equation}
{\bf j}_{r}=m_{r}\int d{\bf v}\,{\bf V}\,f_{r}({\bf v})
\label{2.12}
\end{equation}
is the mass flux for species $r$ relative to the local flow,
\begin{equation}
{\sf P}=\sum_{r}\,\int d{\bf v}\,m_{r}{\bf V}{\bf V}\,f_{r}({\bf  v})
\label{2.13}
\end{equation}
is the total pressure tensor, and
\begin{equation}
{\bf q}=\sum_{r}\,\int d{\bf v}\,\frac{1}{2}m_{r}V^{2}{\bf V}
\,f_{r}({\bf v})
\label{2.14}
\end{equation}
is the total heat flux. The balance equations (\ref{2.9})--(\ref{2.11}) apply regardless of the details of the model for inelastic collisions considered. However, the influence of the collision model appears through the dependence of the cooling rate and the hydrodynamic fluxes on the coefficients of restitution.

As happens for elastic collisions,\cite{GS03} the main advantage of using IMM is
that a velocity moment of order $k$ of the Boltzmann collision operator only involves moments of order less than or equal to $k$. \cite{BC02} This allows one to determine the Boltzmann collisional moments without the explicit knowledge of the velocity distribution function.  The first few moments of the Boltzmann collision operator $J_{rs}[f_r,f_s]$ have been explicitly evaluated in
Appendix \ref{appA}. In particular, according to Eq.\ (\ref{2.7}), the second moment of $J_{rs}[f_r,f_s]$ allows one to get a relationship between the collision frequencies $\omega_{rs}$ and the cooling rates $\zeta_{rs}$. From Eq.\ (\ref{a2}), one easily gets
\begin{equation}
\label{2.15}
\zeta_{rs}=\frac{2\omega_{rs}}{d}\mu_{sr}(1+\alpha _{rs})\left[1-\frac{\mu_{sr}}{2}(1+\alpha_{rs})
\frac{\theta_r+\theta_s}{\theta_s}+\frac{\mu_{sr}(1+\alpha _{rs})-1}{d\rho_sp_r}
{\bf j}_r\cdot {\bf j}_s\right],
\end{equation}
where
\begin{equation}
\label{2.16}
\theta_r=\frac{m_r}{\gamma_r}\sum_{s}m_s^{-1},
\end{equation}
$p_r=n_rT_r$ is the partial pressure of species $r$ and $\gamma_r\equiv T_r/T$ .

In order to get explicit results, one still needs to fix the parameters $\omega_{rs}$.  The most natural choice to optimize the agreement with the IHS results is to adjust the cooling rates $\zeta_{rs}$ for IMM, Eq.\ (\ref{2.15}), to be the same as the ones found for IHS. \cite{GD99} Although the cooling rates are not exactly known for IHS, a good estimate of them can be obtained by considering the local equilibrium approximation for the velocity distribution functions $f_r$, i.e.,
\begin{equation}
\label{2.16bis}
f_r({\bf V})\to n_r\left(\frac{m_r}{2\pi T_r}\right)^{d/2}\exp\left(-\frac{m_rV^2}{2T_r}\right).
\end{equation}
In this approximation, one has \cite{GD99}
\begin{equation}
\label{2.31}
\zeta_{rs}^{\text{IHS}}\to \frac{2\Omega_d}{\sqrt{\pi}d}n_s\mu_{sr}\sigma_{rs}^{d-1}v_0\left(\frac{\theta_r+\theta_s}
{\theta_r\theta_s}\right)^{1/2}(1+\alpha_{rs})\left[1-\frac{\mu_{sr}}{2}(1+\alpha_{rs})
\frac{\theta_r+\theta_s}{\theta_s}\right],
\end{equation}
where $v_0(t)=\sqrt{2T(m_1+m_2)/m_1m_2}$ is a thermal velocity defined in terms of the temperature $T(t)$ of the mixture. Thus, according to Eq.\ (\ref{2.15}), the collision frequencies $\omega_{rs}$ are given by
\begin{equation}
\label{2.32}
\omega_{rs}=4 x_s\left(\frac{\sigma_{rs}}{\sigma_{12}}\right)^{d-1}
\left(\frac{\theta_r+\theta_s}{\theta_r\theta_s}\right)^{1/2}\nu_0,
\end{equation}
where $\nu_0$ is an effective collision frequency given by
\begin{equation}
\label{2.23}
\nu_0=\frac{\Omega_d}{4\sqrt{\pi}}n\sigma_{12}^{d-1}v_0.
\end{equation}
Upon deriving (\ref{2.32}) use has been made of the fact that the mass flux ${\bf j}_r$ vanishes in the local equilibrium approximation (\ref{2.16bis}). In the remainder of this paper, we will take the choice (\ref{2.32}) for $\omega_{rs}$.

\section{Homogeneous cooling state}
\label{sec3}

As a previous step to determine the Navier-Stokes transport coefficients from the Chapman-Enskog method, \cite{CC70} one needs to analyze the {\em homogeneous} solution of the Boltzmann equation (\ref{2.1}). In this case (spatially isotropic homogeneous states), the Boltzmann equation (\ref{2.1}) becomes
\begin{equation}
\label{2.16.1}
\partial_t f_r(v;t)=\sum_s\, J_{rs}[v|f_r,f_s].
\end{equation}
From Eq.\ (\ref{2.16.1}), one has
\begin{equation}
\label{2.17}
\partial_t T=-\zeta T,\quad \partial_t T_r=-\zeta_r T_r.
\end{equation}
The time evolution of the temperature ratio  $\gamma\equiv T_1(t)/T_2(t)$ follows from the second equality of Eq.\ (\ref{2.17}):
 \begin{equation}
\label{2.19}
\partial_t \ln \gamma=\zeta_2- \zeta_1.
\end{equation}
In addition, since the mass flux vanishes in the homogeneous state, Eq.\ (\ref{2.15}) gives the relation
\begin{equation}
\label{2.18bis}
\zeta_r=\sum_{s}\zeta_{rs}=\sum_s\frac{2\omega_{rs}}{d}\mu_{sr}(1+\alpha _{rs})\left[1-\frac{\mu_{sr}}{2}(1+\alpha_{rs})
\frac{\theta_r+\theta_s}{\theta_s}\right],
\end{equation}
where $\omega_{rs}$ is given by Eq.\ (\ref{2.32}).

The so-called homogeneous cooling state (HCS) qualifies as a {\em normal} solution for which all the time dependence of $f_r(v; t)$ is through the global temperature $T(t)$. Consequently, it follows from dimensional analysis that $f_r(v; t)$ has the scaling form
\begin{equation}
\label{2.20}
f_r(v,t)=n_r v_0^{-d}(t)\Phi_r(v/v_0(t)),
\end{equation}
where $v_0$ is the thermal velocity previously introduced. The fact that the time dependence of $f_r(v; t)$ only occurs through the temperature $T(t)$ (which is the relevant one at a hydrodynamic level) implies that the three temperatures $T_1(t)$, $T_2(t)$ and $T(t)$ are proportional to each other and their ratios are all constant. One possibility is that all three temperatures are equal, as in the case of elastic collisions. However, as we will see later, the partial temperatures are different. Since the temperature ratio must be independent of time, Eq.\ (\ref{2.19}) leads to the equality of the cooling rates:
\begin{equation}
\label{2.21}
\zeta_1(t)=\zeta_2(t)=\zeta(t).
\end{equation}
When the expression (\ref{2.18bis}) is substituted into (\ref{2.21}), one gets a closed nonlinear equation  for $\gamma$ whose numerical solution gives the dependence  of the temperature ratio on the parameters of the problem. Except for mechanically equivalent particles,  our results show that although both species have a common cooling rate, their partial temperatures are clearly different. This implies a breakdown of the energy equipartition. The violation of energy equipartition in multicomponent granular systems \cite{GD99,MP99,BT02,MP02,MP02bis,NK02} has also been confirmed in computer simulations \cite{MG02bis,DHGD02,KT03,WJM03} and even observed in real experiments in two \cite{WP02} and three \cite{FM02} dimensions. It must be remarked that the fact that $T_1(t)\neq T_2(t)$ does not mean that there are additional hydrodynamic degrees of freedom since the partial temperatures $T_r$ can be expressed in terms of the granular temperature $T$ as
\begin{equation}
\label{2.21.1}
T_1(t)=\frac{\gamma}{1+x_1(\gamma-1)}T(t),\quad T_2(t)=\frac{1}{1+x_1(\gamma-1)}T(t).
\end{equation}

The problem is therefore to solve the Boltzmann equation for a distribution of the form (\ref{2.20}) subject to the self-consistency constraint (\ref{2.21}). In terms of the reduced velocity $v^*=v/v_0$, the Boltzmann equation (\ref{2.16.1}) for the reduced distribution $\Phi_r$ defined in Eq.\ (\ref{2.20})  becomes
\begin{equation}
\label{2.22}
\frac{1}{2}\zeta^*\frac{\partial}{\partial {\bf v^*}}\cdot \left({\bf v^*}\Phi_r\right)=\sum_sJ_{rs}^*[v^*|\Phi_r,\Phi_s],
\end{equation}
with $\zeta^*=\zeta/\nu_0$.  In addition,
\begin{equation}
\label{2.24}
J_{rs}^*[\Phi_r,\Phi_s]=\frac{\omega_{rs}^*}{\Omega_d}
\int d{\bf v}_{2}^*\int d\widehat{\boldsymbol{\sigma }}\left[ \alpha_{rs}^{-1}\Phi_{r}( v_{1}'^*)\Phi_{s}(
 v_{2}'^*)-\Phi_{r}(v_{1}^*)\Phi_{s}( v_{2}^*)\right],
\end{equation}
where  $\omega_r^*=\omega_r/\nu_0$. Upon writing Eq.\ (\ref{2.22}), we have accounted for the time dependence of $f_r$, which implies that
\begin{equation}
\label{2.25}
\partial_tf_r=-\zeta T\partial_Tf_r=\frac{1}{2}\zeta
\frac{\partial}{\partial {\bf v}}\cdot \left({\bf v}f_r\right).
\end{equation}

Although the exact form of the distribution $\Phi_r$ is not known,  an {\em indirect} information on the behavior of $\Phi_r$ is given through its velocity moments. In particular, the deviation of $\Phi_r$ with respect to its Maxwellian form $\Phi_{r,M}$ can be characterized through the fourth cumulant
\begin{equation}
\label{2.26}
c_r=2\left[\frac{4}{d(d+2)}\theta_r^2\langle v^{*4}\rangle_{r}-1\right],
\end{equation}
where
\begin{equation}
\label{2.27}
\langle v^{*4}\rangle_{r}=\int\,d{\bf v}^*v^{*4}\Phi_r(v^*),
\end{equation}
and
\begin{equation}
\label{2.27.1}
\Phi_{r,M}(v^*)=\pi^{-d/2}\theta_r^{d/2} e^{-\theta_r v^{*2}}.
\end{equation}
To get the fourth velocity moment, we multiply both sides of the Boltzmann equation (\ref{2.22}) by $v^{*4}$ and integrate over the velocity. The result is
\begin{equation}
\label{2.28}
-2\zeta^*\langle v^{*4}\rangle_{r}=\sum_s\int\, d{\bf v}^*v^{*4}J_{rs}^*[v^*|\Phi_r,\Phi_s].
\end{equation}
\begin{figure}
\includegraphics[width=0.25 \columnwidth,angle=-90]{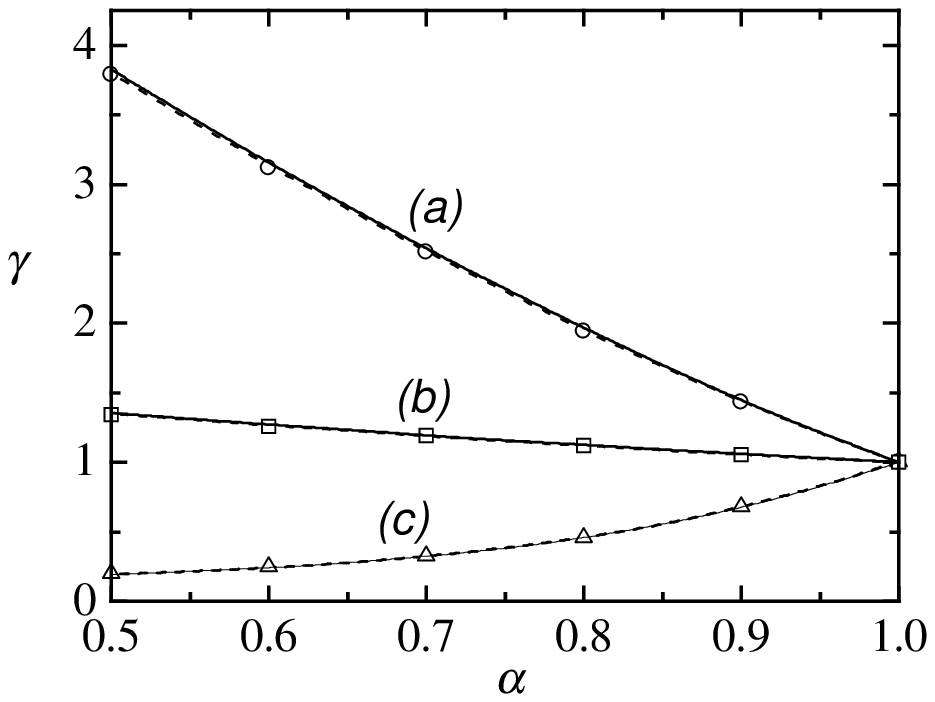}
\caption{Plot of the temperature ratio $\gamma=T_1/T_2$  versus the coefficient of restitution $\alpha$ for $d=3$, $x_1=\case{2}{3}$, $\sigma_1/\sigma_2=1$ and three different values of the mass ratio $ m_1/m_2$: (a) $m_1/m_2=10$ (circles), (b) $m_1/m_2=2$ (squares), and (c) $m_1/m_2=0.1$ (triangles). The solid lines are the results derived here for IMM, the dashed lines correspond to the results obtained for IHS from the first Sonine approximation,\cite{GD99} and the symbols refer to Monte Carlo simulations for IHS. \cite{MG02bis}
\label{fig2}}
\end{figure}
The collisional moment appearing on the right-hand side of (\ref{2.28}) is evaluated in Appendix \ref{appA} with the result
\begin{eqnarray}
\label{2.29}
\int \,d{\bf v}^*v^{*4}J_{rs}^*[\Phi_r,\Phi_s]&=&
\frac{\mu_{sr}(1+\alpha_{rs})}{d(d+2)}\omega_{rs}^*\left\{3\mu_{sr}^3(1+\alpha_{rs})^3\langle v^{*4}\rangle_{s}\right.\nonumber\\
& & +\left[2d+3\mu_{sr}^2(1+\alpha_{rs})^2-6\mu_{sr}(1+\alpha_{rs})+4\right]\left[\mu_{sr}(1+\alpha_{rs})-2\right] \langle v^{*4}\rangle_{r} \nonumber\\
& & +\frac{d(d+2)}{4}\mu_{sr}\theta_r^{-1}\theta_s^{-1}
(1+\alpha_{rs})\left[2d+4-12\mu_{sr}(1+\alpha_{rs})\right.\nonumber\\
& & \left.\left.
+6\mu_{sr}^2(1+\alpha_{rs})^2\right]\right\}.
\end{eqnarray}
Substitution of Eq.\ (\ref{2.29}) into Eq.\ (\ref{2.28}) leads to a coupled set of linear equations for $\langle v^{*4}\rangle_{1}$ and $\langle v^{*4}\rangle_{2}$ (or equivalently, for $c_1$ and $c_2$). The solution of this set gives $c_1$ and $c_2$ in terms of the parameter space of the problem. In the one-dimensional case ($d=1$), our results reduce to the ones previously obtained by Marconi and Puglisi\cite{MP02} from the so-called Maxwell scalar model (i.e., by taking $\omega_{rs}\propto n_s$). Moreover, for mechanically equivalent particles ($\sigma_1=\sigma_2$, $m_1=m_2$, $\alpha_{11}=\alpha_{22}=\alpha_{12}=\alpha$), the results obtained by Santos \cite{S03} for the single gas case are recovered, namely, $\gamma=1$ and
\begin{equation}
\label{2.30}
c_1=c_2=\frac{12(1-\alpha)^2}{4d-7+3\alpha(2-\alpha)}.
\end{equation}

\begin{figure}
\includegraphics[width=0.25 \columnwidth,angle=-90]{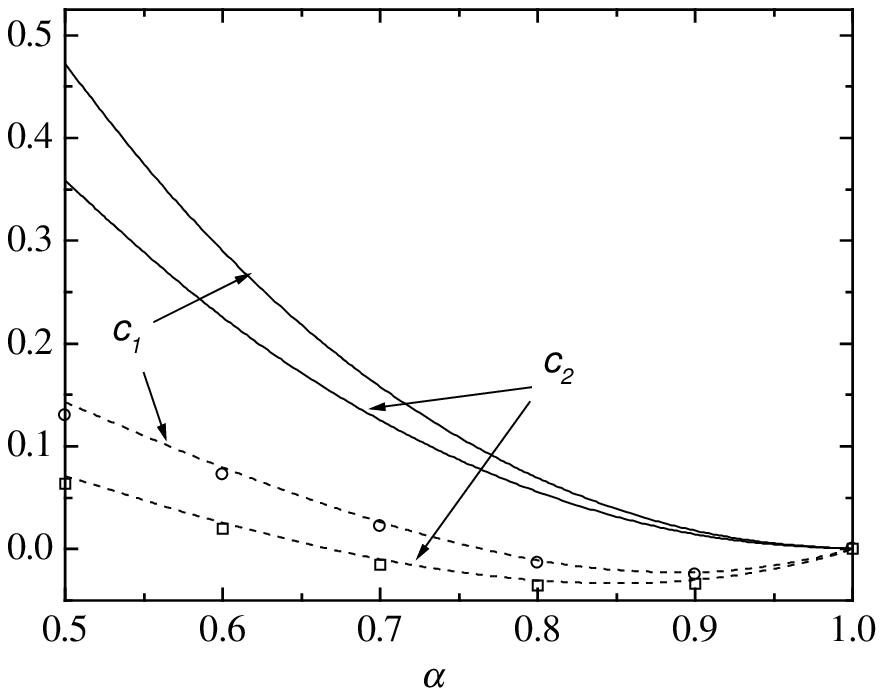}
\caption{Plot of the coefficients $c_r$ versus the coefficient of restitution $\alpha$ for $d=3$,  $x_1=\case{1}{2}$, $\sigma_1/\sigma_2=1$ and $m_1/m_2=2$. The solid lines are the results derived here for IMM, the dashed lines correspond to the results obtained for IHS from the first Sonine approximation, \cite{GD99} and the symbols refer to Monte Carlo simulations for IHS.  \cite{MG02bis}The circles correspond to $c_1$ while the squares correspond to $c_2$.
\label{fig1}}
\end{figure}

A full presentation of the results is difficult due to the many parameters involved in the problem: $\alpha_{11}$, $\alpha_{12}$, $\alpha_{22}$, $m_1/m_2$, $x_1$, and $\sigma_1/\sigma_2$. For the sake of concreteness, henceforth we will consider the case $\alpha_{11}=\alpha_{22}=\alpha_{12}\equiv\alpha$.
In Fig.\  \ref{fig2} we show the dependence of $\gamma$ on $\alpha$ in the three-dimensional case for $x_1=\case{2}{3}$, $\sigma_1/\sigma_2=1$ and three different values of the mass ratio $m_1/m_2$. We include the analytical results obtained for IMM and IHS\cite{GD99} as well as the results (symbols) obtained \cite{MG02bis} by numerically solving the Boltzmann equation by means of the direct simulation Monte Carlo (DSMC) method. \cite{B99} It is apparent that the analytical results for IMM and IHS are practically indistinguishable over the range of values of $\alpha$ considered and that the agreement of both approaches with simulation is excellent. We also observe that the extent of the equipartition violation is greater when the mass disparity is large.  In addition, the temperature of the excess species is larger (smaller) than that of the defect species when the excess species is heavier (lighter) than the defect species. The dependence of the coefficients $c_1$ and $c_2$ on the coefficient of restitution $\alpha$ is plotted in Fig.\ \ref{fig2} in the three-dimensional case for $x_1=\case{1}{2}$, $\sigma_1/\sigma_2=1$ and $m_1/m_2=2$. It can be observed that the HCS of IMM deviates from the Gaussian distribution (which corresponds to $c_1=c_2=0$) much more than the HCS of IHS. This is consistent with the fact that the former models have a stronger overpopulated high energy tail \cite{EB02,EB02bis,MP02} than the latter. \cite{NE98} For both interaction models, the deviation of $\Phi_r$ from its Gaussian form is more significant for the heavy species.  We also see that the value of the kurtosis $c_r$ predicted by the IMM exhibits quantitative discrepancies with the one found for IHS, especially for strong dissipation.

\section{Chapman-Enskog solution of the Boltzmann equation for IMM}
\label{sec4}

In this Section, the Chapman-Enskog method \cite{CC70} generalized to inelastic collisions will be applied to the set of Boltzmann equations (\ref{2.1}) for IMM to get explicit expressions for the Navier-Stokes transport coefficients as functions of the coefficients of restitution and the parameters of the mixture (masses, composition, and sizes).

The balance equations (\ref{2.9})--(\ref{2.11}) become a closed set of hydrodynamic equations for the fields $n_r$, ${\bf u}$, and $T$ once the fluxes (\ref{2.12})--(\ref{2.14}) and the cooling rate $\zeta$ are obtained in terms of the hydrodynamic fields and their gradients. As noted in Sec.\ \ref{sec1}, while the pressure tensor has the same form as for a one-component system, there is greater freedom in representing the heat and mass fluxes.     Here, as done in the IHS case,  \cite{GD02} we take the gradients of the mole fraction $x_1=n_1/n$,
the pressure $p$, the temperature $T$, and the flow velocity ${\bf u}$ as
the relevant ones.  Thus, in this representation, the phenomenological constitutive
relations for the fluxes in the low-density regime have the forms \cite{GM84}
\begin{equation}
\label{3.1}
{\bf j}_1=-\frac{m_1m_2n}{\rho}D\nabla x_1-\frac{\rho}{p}D_p\nabla p-
\frac{\rho}{T}D'\nabla T,\quad {\bf j}_2=-{\bf j}_1,
\end{equation}
\begin{equation}
\label{3.2}
{\bf q}=-T^2D''\nabla x_1-L\nabla p-\lambda\nabla T,
\end{equation}
\begin{equation}
\label{3.3}
P_{ij}=p\delta_{ij}-\eta\left(\nabla_j u_i+
\nabla_i u_j-\frac{2}{d}\delta_{ij}\nabla \cdot {\bf u}\right).
\end{equation}
The transport coefficients in these equations are the diffusion coefficient $D$, the thermal
diffusion coefficient $D'$, the pressure diffusion coefficient $D_p$, the
Dufour coefficient $D''$, the thermal conductivity $\lambda$, the pressure
energy coefficient $L$, and the shear viscosity $\eta$.

The Chapman-Enskog method assumes the existence of a {\em normal} solution
in which all space and time dependence of the distribution function occurs
through a functional dependence on the hydrodynamic fields
\begin{equation}
\label{3.4}
f_r({\bf r}, {\bf v},t)=f_r[{\bf v}|x_1(t), p(t), T(t), {\bf u}(t)].
\end{equation}
This functional dependence can be made local in space and time by means of
an expansion in gradients of the fields. Thus, we write $f_{r}$ as a series
expansion in a formal parameter $\epsilon $ measuring the nonuniformity of
the system,
\begin{equation}
f_{r}=f_{r}^{(0)}+\epsilon \,f_{r}^{(1)}+\epsilon^2 \,f_{r}^{(2)}+\cdots \;,
\label{3.5}
\end{equation}
where each factor of $\epsilon $ means an implicit gradient of a
hydrodynamic field. The local reference state $f_r^{(0)}$ is chosen to give
the same first moments as the exact distribution $f_r$, or equivalently, the remainder of the expansion must obey the orthogonality conditions
\begin{equation}
\label{3.5.1}
\int\, d{\bf v} \left[f_r({\bf v})-f_r^{(0)}({\bf v})\right]=0,
\end{equation}
\begin{equation}
\label{3.5.2}
\sum_r\, \int\, d{\bf v} m_r{\bf v}\,\left[f_r({\bf v})-f_r^{(0)}({\bf v})\right]=0,
\end{equation}
\begin{equation}
\label{3.5.3}
\sum_r\, \int\, d{\bf v} \frac{m_r}{2} v^2\,\left[f_r({\bf v})-f_r^{(0)}({\bf v})\right]=0.
\end{equation}
Use of the expansion (\ref{3.5}) in the definitions of the fluxes (\ref{2.12})--(\ref{2.14}) and the cooling rate (\ref{2.7}) gives the corresponding expansion for these quantities.  The time derivatives
of the fields are also expanded as $\partial_t=\partial_t^{(0)}+\epsilon
\partial_t^{(1)}+\cdots$. The coefficients of the time derivative expansion
are identified from the balance equations (\ref{2.9})--(\ref{2.11}) after
expanding the fluxes and the cooling rate $\zeta$. In particular, the macroscopic balance equations to zeroth order become
\begin{equation}
\label{3.6}
\partial_t^{(0)}x_r=0,\quad \partial_t^{(0)}{\bf u}={\bf 0},\quad T^{-1}\partial_t^{(0)}T=p^{-1}\partial_t^{(0)}p=-\zeta^{(0)}.
\end{equation}
Here, we have taken into account that in the Boltzmann equation (\ref{2.1}) the effective collision frequency $\omega_{rs}\propto n_s T^{1/2}\propto x_s p T^{-1/2}$ is assumed to be a functional of $f_r$ and $f_s$ only through the mole fraction $x_s$, the pressure $p$, and the temperature $T$. As a consequence, $\omega_{rs}^{(0)}=\omega_{rs}$, and $\omega_{rs}^{(1)}=\omega_{rs}^{(2)}=\cdots =0$.

In the zeroth order, $f_r^{(0)}$ obeys the kinetic equation
\begin{equation}
\label{3.7}
\frac{1}{2}\zeta^{(0)}\frac{\partial}{\partial {\bf V}}
\cdot \left({\bf V}f_r^{(0)}\right)=\sum_{s}J_{rs}[f_r^{(0)},f_s^{(0)}],
\end{equation}
where use has been made of the relation (\ref{2.25}) with $\zeta^{(0)}=\zeta_r^{(0)}$ given by Eq.\ (\ref{2.18bis}).  The distribution $f_r^{(0)}$ is given by the scaling form  (\ref{2.20}) except that $n_r\to n_r({\bf r},t)$ and $T\to T({\bf r},t)$ are local quantities and ${\bf v}\to {\bf V}={\bf v}-{\bf u}({\bf r},t)$. Since $f_r^{(0)}$ is isotropic, it follows that
\begin{equation}
\label{3.9}
{\bf j}_1^{(0)}={\bf 0},\quad{\bf q}^{(0)}={\bf 0},\quad P_{ij}^{(0)}=p\delta_{ij},
\end{equation}
where $p=nT$ is the hydrostatic pressure.

In the first order, the distribution function $f_r^{(1)}$ verifies the kinetic equation
\begin{equation}
\label{3.10}
\left(\partial_t^{(0)}+{\cal L}_r\right)f_r^{(1)}+{\cal M}_rf_s^{(1)}=
-\left(D_t^{(1)}+{\bf V}\cdot \nabla\right)f_r^{(0)},
\end{equation}
where it is understood that $r\neq s$. Here, $D_t^{(1)}=\partial_t^{(1)}+{\bf u}\cdot \nabla$ and we have introduced the linearized collision operators
\begin{equation}
\label{3.11}
{\cal L}_rf_r^{(1)}=-\left(J_{rr}[f_r^{(0)},f_r^{(1)}]+
J_{rr}[f_r^{(1)},f_r^{(0)}]+J_{rs}[f_r^{(1)},f_s^{(0)}]\right),
\end{equation}
\begin{equation}
\label{3.12}
{\cal M}_rf_s^{(1)}=-J_{rs}[f_r^{(0)},f_s^{(1)}].
\end{equation}
The action of the material time derivatives $D_t^{(1)}$ on the hydrodynamic fields is
\begin{equation}
\label{3.13}
D_t^{(1)}x_1=0,\quad \frac{d}{d+2}D_t^{(1)}\ln p=\frac{d}{2}D_t^{(1)}\ln T=
-\nabla \cdot {\bf u},\quad D_t^{(1)}{\bf u}=-\rho^{-1}\nabla p.
\end{equation}
In these equations use has been made of the results (\ref{3.9}) and the identity $\zeta^{(1)}=0$. The last equality easily follows from Eq.\ (\ref{2.15}).  The right-hand side of Eq.\ (\ref{3.10}) can be evaluated by using Eqs.\ (\ref{3.13}) and so the equation for $f_r^{(1)}$ can be written as
 \begin{equation}
\label{3.14}
\left(\partial_t^{(0)}+{\cal L}_r\right)f_r^{(1)}+{\cal M}_rf_s^{(1)}=
{\bf A}_r\cdot \nabla x_1+
{\bf B}_r\cdot \nabla p+{\bf C}_r\cdot \nabla
T+D_{r,ij}\nabla_{i}u_{j},
\end{equation}
where
\begin{equation}
{\bf A}_{r}({\bf V})=-\left( \frac{\partial }{\partial x_{1}}
f_{r}^{(0)}\right) _{p,T}{\bf V},  \label{3.15}
\end{equation}
\begin{equation}
{\bf B}_{r}({\bf V})=-\frac{1}{p}\left[ f_{r}^{(0)}{\bf V}+\frac{p}{\rho }
\left( \frac{\partial }{\partial {\bf V}}f_{r}^{(0)}\right) \right],
\label{3.16}
\end{equation}
\begin{equation}
{\bf C}_{r}({\bf V})=\frac{1}{T}\left[ f_{r}^{(0)}+\frac{1}{2}\frac{
\partial }{\partial {\bf V}}\cdot \left( {\bf V}f_{r}^{(0)}\right) \right]
{\bf V},
\label{3.17}
\end{equation}
\begin{equation}
D_{r,ij }({\bf V})=\frac{\partial }{\partial V_{j}}\left(V_i
f_{r}^{(0)}\right)-\frac{1}{d}\delta _{ij} \frac{\partial }{
\partial {\bf V}}\cdot\left({\bf V}f_{r}^{(0)}\right).
\label{3.18}
\end{equation}
It is worth noting that Eqs.\ (\ref{3.14})--(\ref{3.18}) have the same structure as that of the Boltzmann equation for IHS. \cite{GD02} The only difference between both models lies in the explicit form of the linearized operators ${\cal L}_r$ and ${\cal M}_r$.

Now we are in conditions to get the expressions for the mass flux, the pressure tensor, and the heat flux in the first order of gradients. These expressions allows one to identify the relevant transport coefficients of the mixture through Eqs.\ (\ref{3.1})--(\ref{3.3}).

\section{Navier-Stokes transport coefficients}
\label{sec5}

This Section is devoted to the determination of the Navier-Stokes transport coefficients associated with the irreversible fluxes. We only display here the final expressions for the transport coefficients with technical details given in Appendix \ref{appB}.

To first order in the hydrodynamic gradients, the mass flux has the form given by Eq.\ (\ref{3.1}). The transport coefficients $D$, $D_p$, and $D'$ are given by
\begin{equation}
D=\frac{\rho }{m_{1}m_{2}n}\left( \nu -\frac{1}{2}\zeta ^{(0)}\right) ^{-1}
\left[ p\left( \frac{\partial }{\partial x_{1}}x_{1}\gamma_{1}\right) _{p,T}+\rho
\left( \frac{\partial \zeta ^{(0)}}{\partial x_{1}}\right) _{p,T}\left(
D_{p}+D^{\prime }\right) \right] ,  \label{4.8}
\end{equation}
\begin{equation}
D_{p}=\frac{n_{1}T_{1}}{\rho }\left( 1-\frac{m_{1}nT}{\rho T_{1}}\right)
\left( \nu -\frac{3}{2}\zeta ^{(0)}+\frac{\zeta ^{(0)2}}{2\nu }\right) ^{-1},
\label{4.9}
\end{equation}
\begin{equation}
D^{\prime }=-\frac{\zeta ^{(0)}}{2\nu }D_{p},  \label{4.10}
\end{equation}
where
\begin{equation}
\label{4.3}
\nu=\frac{\rho\omega_{12}}{d\rho_2}\mu_{21}(1+\alpha_{12})=\frac{4}{d}\frac{\rho}{n(m_1+m_2)}
\left(\frac{\theta_1+\theta_2}{\theta_1\theta_2}\right)^{1/2}\nu_0(1+\alpha_{12}).
\end{equation}
Since ${\bf j}_{1}=-{\bf j}_{2}$ and $\nabla x_{1}=-\nabla x_{2}$, $D$ should be symmetric while $D_{p}$ and $D'$ should be antisymmetric with respect to the exchange $1\leftrightarrow 2$ . This can be easily verified by noting that $n_{1}T_{1}+n_{2}T_{2}=nT$.  For elastic collisions, $\zeta^{(0)}=0$, $T_1=T_2=T$, and so Eqs.\ (\ref{4.8})--(\ref{4.10}) become
\begin{equation}
\label{4.11}
D=\frac{d}{8}\frac{m_1+m_2}{m_1m_2}\frac{p}{\nu_0},
\end{equation}
\begin{equation}
\label{4.11bis}
D_p=\frac{d}{8}(m_2^2-m_1^2)\frac{n_1n_2p}{\rho^3\nu_0},\quad D'=0,
\end{equation}
where use has been made again of Eq.\ (\ref{2.32}). In the case of mechanically equivalent particles, $\gamma=1$, $D_p=D'=0$, so that Eq.\ (\ref{4.8}) gives the expression of the self-diffusion coefficient
\begin{equation}
\label{4.11bis.1}
D=\frac{p}{m}\left(\nu-\frac{1}{2}\zeta^{(0)}\right)^{-1}.
\end{equation}

The pressure tensor has the form (\ref{3.3}) with the shear viscosity coefficient $\eta$ given by
\begin{equation}
\label{4.19}
\eta=\eta_1+\eta_2.
\end{equation}
The expression of the partial contributions $\eta_r$ is
\begin{equation}
\label{4.22}
\eta_1=2\frac{p_1(2\tau_{22}-\zeta^{(0)})-2p_2\tau_{12}}{\zeta^{(0)2}-2\zeta^{(0)}
(\tau_{11}+\tau_{22})+4(\tau_{11}\tau_{22}-\tau_{12}\tau_{21})},
\end{equation}
where the quantities $\tau_{11}$ and $\tau_{12}$ are defined by Eqs.\ (\ref{4.14}) and (\ref{4.15}), respectively. A similar expression can be obtained for $\eta_2$ by just making the changes $1 \leftrightarrow 2$. For mechanically equivalent particles, Eq.\ (\ref{4.22}) yields $\eta_1/x_1=\eta_2/x_2=\eta$, where
\begin{equation}
\label{4.22.1}
\eta=\frac{p}{\nu_{\eta}-\case{1}{2}\zeta^{(0)}},\quad \nu_{\eta}=\frac{(1+\alpha)(d+1-\alpha)}{d(d+2)}\omega,
\end{equation}
and $\omega=\omega_{rs}/x_s$. The expression (\ref{4.22.1}) coincides with the one previously derived in the single gas case. \cite{S03}

The case of the heat flux is more involved. Its form is given by Eq.\ (\ref{3.2}) where the coefficients $D''$, $L$ and $\lambda$ are
\begin{equation}
\label{4.34}
D''=D_1''+D_2'', \quad L=L_1+L_2,\quad \lambda=\lambda_1+\lambda_2.
\end{equation}
By using matrix notation, the coupled set of six equations for the unknowns
\begin{equation}
\label{4bis.1}
\{D_1'', D_2'', L_1, L_2, \lambda_1, \lambda_2\}
\end{equation}
can be written as
\begin{equation}
\label{4.36}
\Lambda_{\sigma \sigma'}X_{\sigma'}=Y_{\sigma}.
\end{equation}
Here, $X_{\sigma'}$ is the column matrix defined by the set (\ref{4bis.1}) and $\Lambda_{\sigma\sigma'}$ is the square matrix
\begin{equation}
\label{4.38}
\Lambda=\left(
\begin{array} {cccccc}
T^2(\frac{3}{2}\zeta^{(0)}-\beta_{11})& -T^2\beta_{12}&
p\left(\frac{\partial \zeta ^{(0)}}{\partial x_{1}}\right)_{p,T}&0&
T\left( \frac{\partial \zeta ^{(0)}}{\partial x_{1}}\right)_{p,T}&0 \\
-T^2\beta_{21}&T^2(\frac{3}{2}\zeta^{(0)}-\beta_{22})&0&
p\left( \frac{\partial \zeta ^{(0)}}{\partial x_{1}}\right)_{p,T}&0&
T\left( \frac{\partial \zeta ^{(0)}}{\partial x_{1}}\right)_{p,T}\\
0& 0& \frac{5}{2}\zeta^{(0)}-\beta_{11}& -\beta_{12}&
T\zeta^{(0)}/p&0\\
0& 0 &- \beta_{21} & \frac{5}{2}\zeta^{(0)}-\beta_{22}& 0 &
T\zeta^{(0)}/p\\
0& 0&-p\zeta^{(0)}/2T&0&\zeta^{(0)}-\beta_{11}&- \beta_{12}\\
0& 0& 0&-p\zeta^{(0)}/2T&-\beta_{21}&\zeta^{(0)}-\beta_{22}
\end{array}
\right).
\end{equation}
The column matrix ${\bf Y}$ is
\begin{equation}
\label{4.39}
{\bf Y}=\left(
\begin{array}{c}
Y_1\\
Y_2\\
Y_3\\
Y_4\\
Y_5\\
Y_6
\end{array}
\right),
\end{equation}
where
\begin{equation}
\label{4.40}
Y_1=\frac{m_1m_2n}{\rho}A_{12}D-
\frac{d+2}{2}\frac{nT^2}{m_1}\frac{\partial}{\partial x_1}\left[\left(1+\frac{c_1}{2}\right)x_1\gamma_1^2\right],
\end{equation}
\begin{equation}
\label{4.40.1}
Y_2=-\frac{m_1m_2n}{\rho}A_{21}D-
\frac{d+2}{2}\frac{nT^2}{m_2}\frac{\partial}{\partial x_1}\left[\left(1+\frac{c_2}{2}\right)x_2\gamma_2^2\right],
\end{equation}
\begin{equation}
\label{4.41}
Y_3=\frac{\rho}{p}A_{12}D_p-\frac{d+2}{2}
\frac{n_1T_1^2}{m_1p}
\left(1-\frac{m_1p}{\rho T_1}+\frac{c_1}{2}\right),
\end{equation}
\begin{equation}
\label{4.41.1}
Y_4=-\frac{\rho}{p}A_{21}D_p -\frac{d+2}{2}
\frac{n_2T_2^2}{m_2p}
\left(1-\frac{m_2p}{\rho T_2}+\frac{c_2}{2}\right),
\end{equation}
\begin{equation}
\label{4.42}
Y_5=\frac{\rho}{T}A_{12}D' -\frac{d+2}{2}\frac{n_1T_1^2}{m_1T}\left(1+\frac{c_1}{2}\right),
\end{equation}
\begin{equation}
\label{4.42.1}
Y_6=-\frac{\rho}{T}A_{21}D' -\frac{d+2}{2}\frac{n_2T_2^2}{m_2T}\left(1+\frac{c_2}{2}\right).
\end{equation}
The expressions of the quantities $\beta_{rs}$ and $A_{rs}$ are given in Appendix \ref{appB}.

The solution to Eq.\ (\ref{4.36}) is
\begin{equation}
\label{4.43}
X_{\sigma}=\left(\Lambda^{-1}\right)_{\sigma \sigma'}Y_{\sigma'}.
\end{equation}
This relation provides an explicit expression for the coefficients
$D_r''$, $L_r$ and $\lambda_r$ in terms of the coefficients of restitution and the parameters of the mixture. From these expressions one easily gets the transport coefficients $D''$, $L$, and $\lambda$ from Eq.\ (\ref{4.34}). As expected, Eqs.\ (\ref{4.36})--(\ref{4.42.1}) show that $D''$ is antisymmetric with respect to the change $1\leftrightarrow 2$, while $L$ and $\lambda$ are symmetric. This implies that in the case of mechanically equivalent particles, the Dufour coefficient $D''$ vanishes. Furthermore, in this limit, Eq.\ (\ref{4.43}) leads to the following expression for the heat flux
\begin{equation}
\label{4.43.1}
{\bf q}=-\kappa\nabla T-\mu \nabla n,
\end{equation}
where
\begin{equation}
\label{4.43.2}
\kappa=\lambda+n L=\frac{d+2}{2}\frac{p}{m}\frac{1+c}{\nu_{\kappa}-2\zeta^{(0)}},
\end{equation}
\begin{equation}
\label{4.43.3}
\mu=T L=\frac{T}{n}\frac{\kappa(\zeta^{(0)}+\case{1}{2}c\nu_{\kappa})}{(1+c)(\nu_{\kappa}-\case{3}{2}\zeta^{(0)})},
\end{equation}
with
\begin{equation}
\label{4.43.4}
\nu_{\kappa}=\frac{4(d-1)+(8+d)(1-\alpha)}{4d+4(1-\alpha)}\nu_{\eta}.
\end{equation}
Note that upon writing Eq.\ (\ref{4.43.1}) use has been made of the relation $\nabla p=n\nabla T+T\nabla n$. Again, Eqs.\ (\ref{4.43.1})--(\ref{4.43.4}) coincide with results derived for a single gas.  \cite{S03} Using Eqs.\ (\ref{4.43.2})--(\ref{4.43.4}), it can be seen that the coefficients $\kappa$ and $\mu$ diverge for $d=2$ and $d=3$ when $\alpha=\alpha_0=(4-d)/3d$ and so, both coefficients become negative for $0\leq \alpha < \alpha_0$. This  unphysical behavior could be due to the failure of the hydrodynamic description for values of the coefficient of restitution smaller than $\alpha_0$ or perhaps to the existence of a certain type of hydrodynamic instability. However, given that the value of $\alpha_0$ is quite small ($\alpha_0=\frac{1}{3}$ at $d=2$ and $\alpha_0=\frac{1}{9}$ at $d=3$) this singular behavior could be interpreted as a hydrodynamic breakdown indeed. Elucidation of this point requires further analysis. It must be remarked that the above drawback is absent for IHS since all the transport coefficients are regular functions of $\alpha$ for all $d$.

\begin{figure}
\includegraphics[width=0.25 \columnwidth,angle=-90]{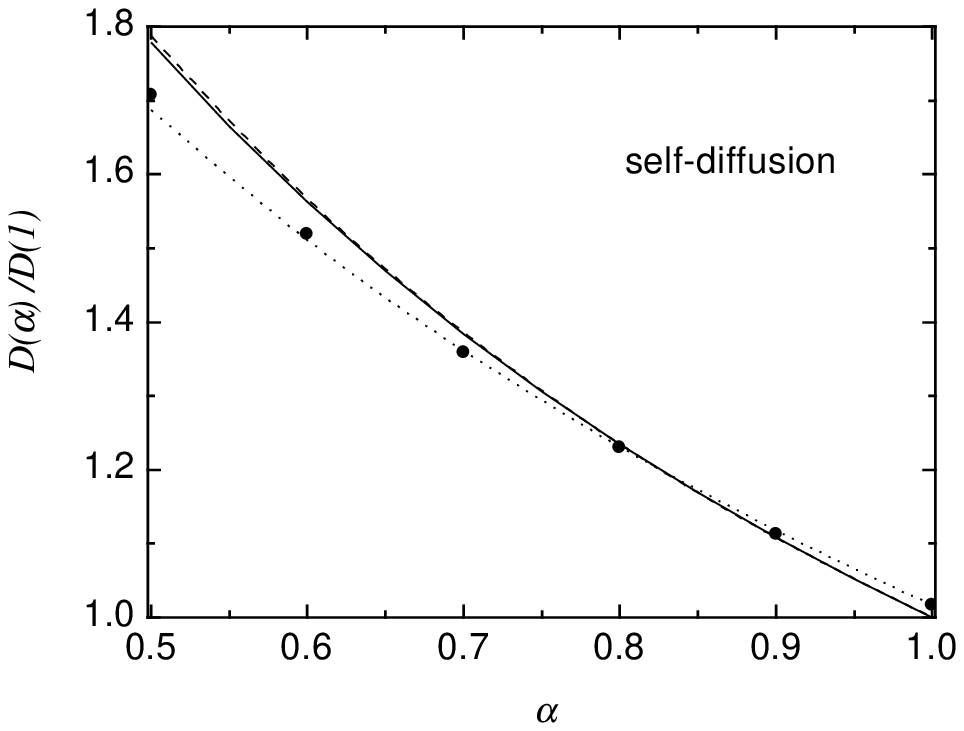}
\caption{Plot of the reduced self-diffusion coefficient $D(\alpha)/D(1)$ as a function of the coefficient of restitution $\alpha$ in the three-dimensional case for IMM (solid line) and for IHS in the first Sonine approximation (dashed line) \cite{BMCG00} and in the second Sonine approximation (dotted line). \cite{GM04} The symbols refer to Monte Carlo simulations for IHS.\cite{GM04,BMCG00}
\label{fig3}}
\end{figure}

\begin{figure}
\includegraphics[width=0.25 \columnwidth,angle=-90]{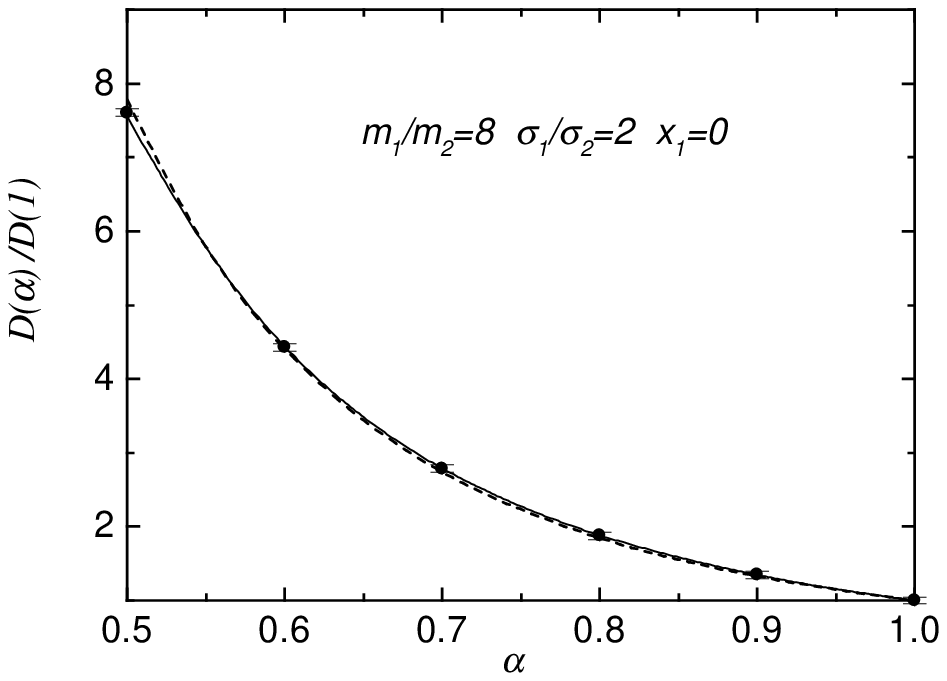}
\caption{Plot of the reduced diffusion coefficient $D(\alpha)/D(1)$ as a function of the coefficient of restitution $\alpha$ in the three-dimensional case for $\sigma_1/\sigma_2=2$, $x_1=0$ (tracer limit) and $m_1/m_2=8$. The solid line corresponds to the exact results obtained here for IMM while the dashed line is the result derived for IHS in the second Sonine approximation.\cite{GM04} The symbols refer to Monte Carlo simulations for IHS.\cite{GM04}
\label{fig4}}
\end{figure}

\begin{figure}
\includegraphics[width=0.25 \columnwidth,angle=-90]{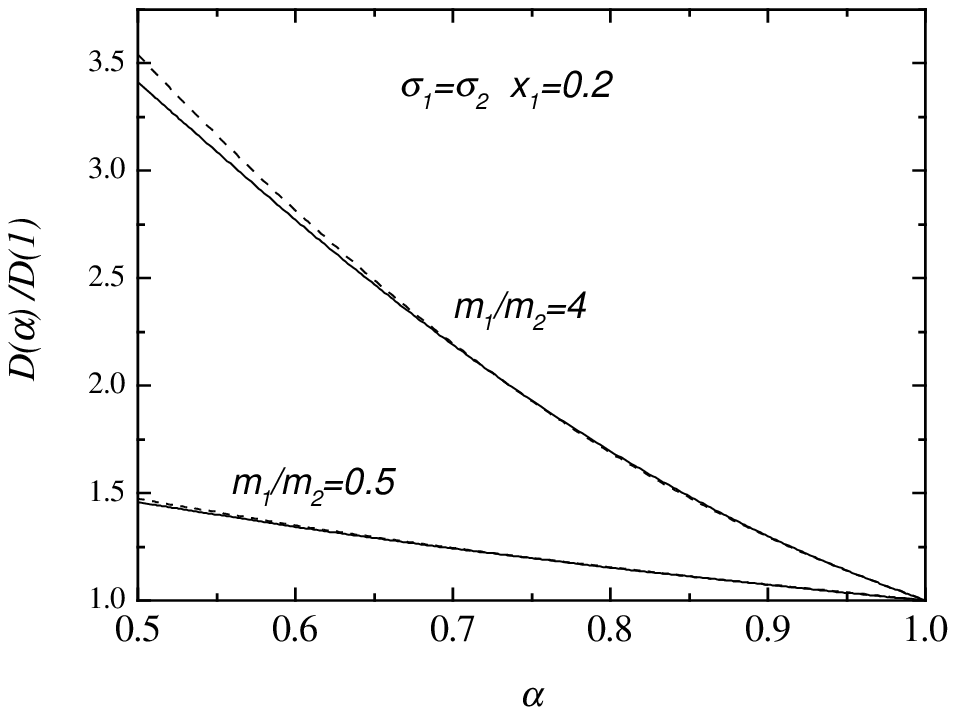}
\caption{Plot of the reduced diffusion coefficient $D(\alpha)/D(1)$ as a function of the coefficient of restitution $\alpha$ in the three-dimensional case for $\sigma_1=\sigma_2$, $x_1=0.2$, and two different values of the mass ratio: $m_1/m_2=0.5$ and $m_1/m_2=4$. The solid lines correspond to the exact results obtained here for IMM while the dashed lines are the results derived for IHS in the first Sonine approximation. \cite{GD02}
\label{fig7}}
\end{figure}

\section{Comparison with the transport coefficients for IHS}
\label{sec6}

The expressions for the transport coefficients of IHS described by the Boltzmann equation have been  obtained by Garz\'o and Dufty \cite{GD02} in the leading Sonine approximation for a three-dimensional system. These expressions have been then evaluated for a variety of mass and diameter ratios in the cases of the diffusion coefficient \cite{GM04} and the shear viscosity coefficient, \cite{MG03bis} showing quite a good agreement with Monte Carlo simulations. In this section, we compare the results derived here for IMM for the transport coefficients entering in the mass and momentum fluxes, namely, $D, D_p, D'$, and $\eta$ with those obtained for IHS in the case $d=3$. As in the cases studied in Section \ref{sec3}, we consider for the sake of simplicity a common coefficient of restitution, i.e.,  $\alpha_{11}=\alpha_{12}=\alpha_{22}\equiv \alpha$.

\begin{figure}
\includegraphics[width=0.25 \columnwidth,angle=-90]{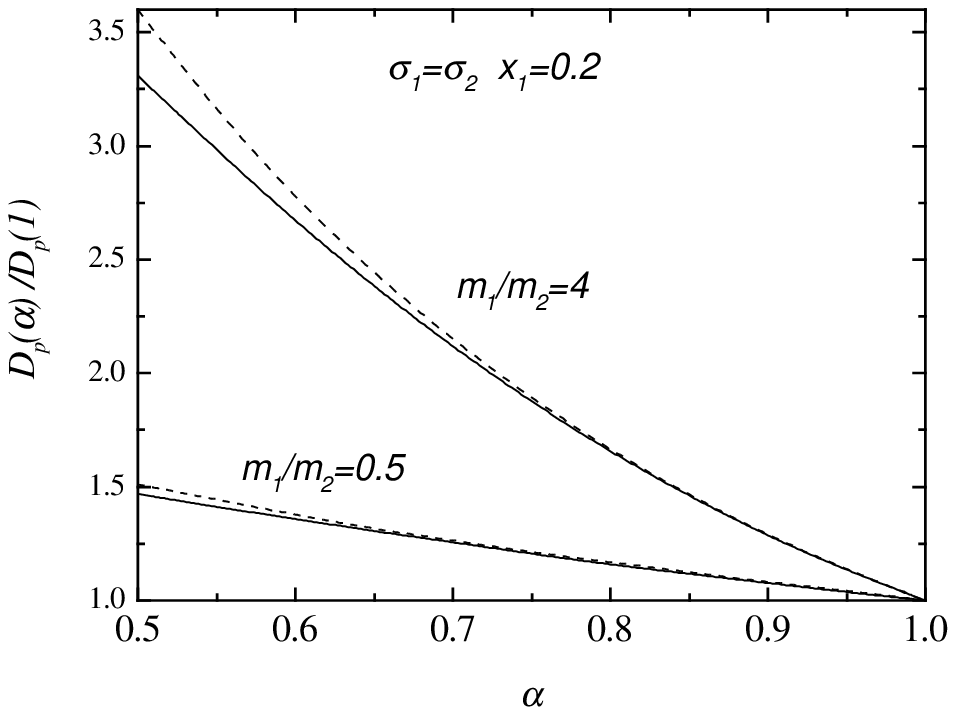}
\caption{Plot of the reduced pressure diffusion coefficient $D_p(\alpha)/D_p(1)$ as a function of the coefficient of restitution $\alpha$ in the three-dimensional case for $\sigma_1=\sigma_2$, $x_1=0.2$, and two different values of the mass ratio: $m_1/m_2=0.5$ and $m_1/m_2=4$. The solid lines correspond to the exact results obtained here for IMM while the dashed lines are the results derived for IHS in the first Sonine approximation. \cite{GD02}
\label{fig5}}
\end{figure}

Let us consider first, the diffusion coefficient $D$ in the limit cases of self-diffusion (mechanically equivalent particles) and tracer concentration ($x_1\to 0$). Figure\ \ref{fig3} shows  the reduced self-diffusion coefficient $D(\alpha)/D(1)$  as a function of the coefficient of restitution $\alpha$. Here, $D(1)$ refers to the self-diffusion coefficient (\ref{4.11bis.1}) for elastic collisions. We include the results obtained for IHS by using the first Sonine approximation (dashed line), \cite{DG01,BMCG00} the second Sonine approximation (dotted line) and by Monte Carlo simulations (symbols). \cite{GM04} We observe that the agreement between the predictions of the first Sonine approximation for IHS and for IMM is excellent in the whole range of values of $\alpha$ analyzed. Moreover, both theories compare quite well with computer simulations even beyond the quasielastic limit (say for instance, $\alpha \geq 0.8$). However, as dissipation increases, the agreement between theory and simulation is improved when one considers the second Sonine approximation. For mechanically different particles, in Fig.\ \ref{fig4} we plot $D(\alpha)/D(1)$ versus $\alpha$ in the tracer limit ($x_1\to 0$) for $m_1/m_2=8$ and $\sigma_1/\sigma_2=2$ as given by the exact results for IMM, the second Sonine approximation for IHS \cite{GM04} and by Monte Carlo simulations. We see that both theories are practically indistinguishable and present an excellent agreement with simulation data.

\begin{figure}
\includegraphics[width=0.25 \columnwidth,angle=-90]{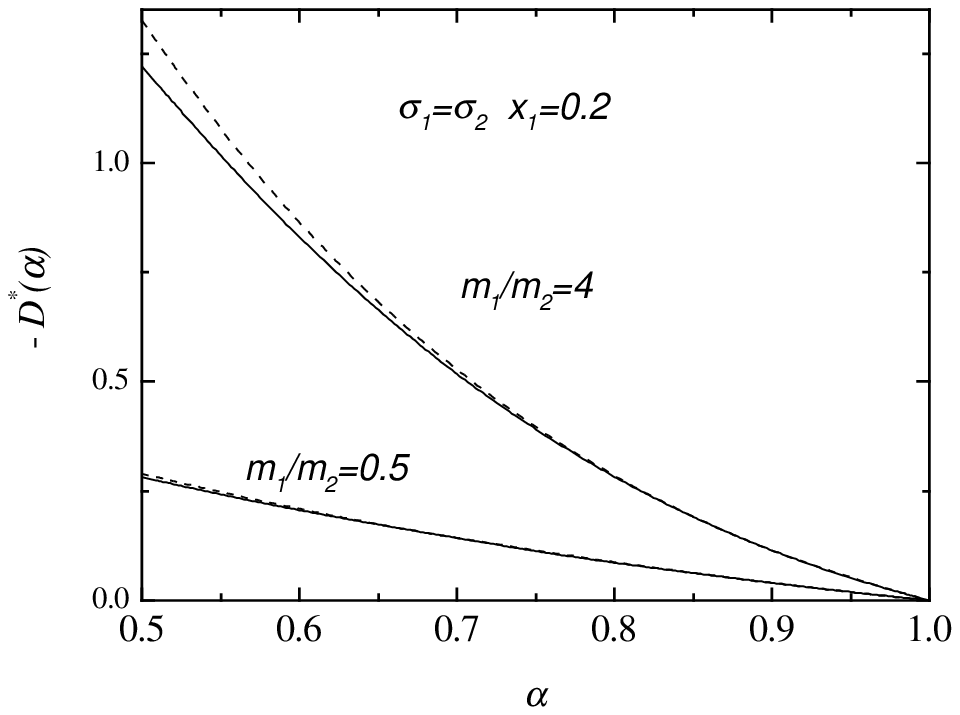}
\caption{Plot of the reduced coefficient $-D^*(\alpha)=-D'(\alpha)/D_p(1)$ as a function of the coefficient of restitution $\alpha$ in the three-dimensional case for $\sigma_1=\sigma_2$, $x_1=0.2$, and two different values of the mass ratio: $m_1/m_2=0.5$ and $m_1/m_2=4$. The solid lines correspond to the exact results obtained here for IMM while the dashed lines are the results derived for IHS in the first Sonine approximation. \cite{GD02}
\label{fig6}}
\end{figure}

Beyond the above two special cases, Figs.\ \ref{fig7}--\ref{fig5} show the dependence of the reduced coefficients $D(\alpha)/D(1)$, $D_p(\alpha)/D_p(1)$, and $-D^*(\alpha)$ on the coefficient of restitution for $\sigma_1=\sigma_2$, $x_1=0.2$, and two values of the mass ratio ($m_1/m_2=0.5$ and 4). Here, $D_p(1)$ is given by the first equality of Eq.\ (\ref{4.11bis}) and $D^*(\alpha)=-(\zeta^{(0)}/2\nu)D_p(\alpha)/D_p(1)=-D'(\alpha)/D_p(1)$. The solid lines correspond to the IMM results while the dashed lines refer to the results obtained for IHS in the first Sonine approximation. \cite{GD02} We observe that, in general, the qualitative behavior of IHS is well captured by the IMM. At a quantitative level, for not strong dissipation (say for instance, $\alpha \geq 0.8$) the agreement between the results derived for IMM and IHS is again quite good, especially in the case of $m_1/m_2=0.5$. Nevertheless, the discrepancies between both interaction models increase as the coefficient of restitution decreases. For instance, at $\alpha=0.5$, the discrepancies for $D(\alpha)/D(1)$, $D_p(\alpha)/D_p(1)$, and $-D^*(\alpha)$  are about  4\%, 5\%, and 8\%, respectively, for $m_1/m_2=4$, while they are about 1\%, 3\%, and 3\%, respectively, for $m_1/m_2=0.5$.

\begin{figure}
\includegraphics[width=0.25 \columnwidth,angle=-90]{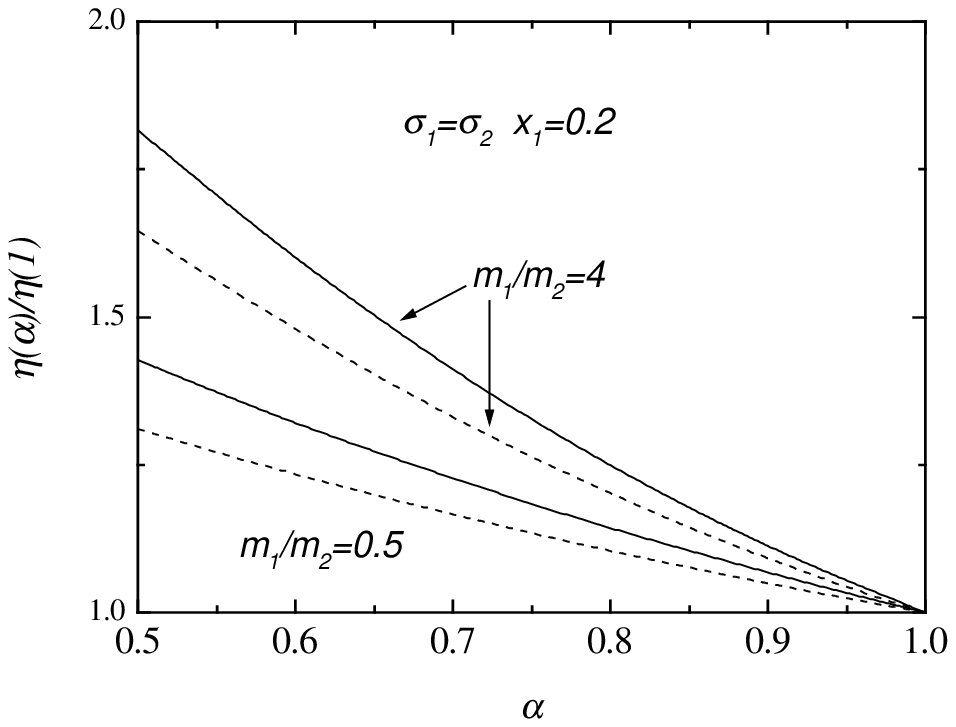}
\caption{Plot of the reduced shear viscosity coefficient $\eta(\alpha)/\eta(1)$ as a function of the coefficient of restitution $\alpha$ in the three-dimensional case for $\sigma_1=\sigma_2$, $x_1=0.2$, and two different values of the mass ratio: $m_1/m_2=0.5$ and $m_1/m_2=4$. The solid lines correspond to the exact results obtained here for IMM while the dashed lines are the results derived for IHS in the first Sonine approximation. \cite{MG03bis}
\label{fig8}}
\end{figure}

The dependence of the reduced shear viscosity coefficient $\eta(\alpha)/\eta(1)$ on dissipation is plotted in Fig.\ \ref{fig8} for $\sigma_1=\sigma_2$, $x_1=0.2$, and for two values of the mass ratio ($m_1/m_2=0.5$ and 4). Here, $\eta(1)$ is the corresponding value of the shear viscosity coefficient for elastic collisions. Although again the IMM predictions compare qualitatively well with the ones derived for IHS, we observe that in general the discrepancies for both interaction models at the level of the shear viscosity are larger than those found for the transport coefficients of the mass flux.  The discrepancies observed in Fig.\ \ref{fig8} are similar to those found in the single gas case. \cite{S03}

\section{Discussion}
\label{sec7}

The primary objective of this work has been to derive the hydrodynamic equations of a granular binary mixture from the Boltzmann kinetic theory for {\em inelastic} Maxwell models (IMM). In the Boltzmann equation for inelastic Maxwell models (IMM), the collision rate of inelastic hard spheres (IHS) is replaced by an effective collision rate independent of the relative velocity of the two colliding particles. As in the elastic case,\cite{GS03} this property allows us to get exactly the velocity moments of the Boltzmann collision integrals without the explicit knowledge of the velocity distribution function. Here, the Chapman-Enskog method has been applied to get a {\em normal} solution of the Boltzmann equation for states near the local homogenous cooling state. The derivation of the Navier-Stokes hydrodynamic equations consists of two steps. As a first step, the reference homogenous cooling state for a mixture of inelastic Maxwell gases is analyzed to provide the proper basis for description of transport due to spatial inhomogeneities. As in the case of inelastic hard spheres (IHS), \cite{GD99} our solution for the homogeneous state shows that the kinetic temperatures for each species are clearly different so that the total energy is not equally distributed between both species (breakdown of energy equipartition). In addition, we also compute  the fourth cumulant of the velocity distribution functions, which is a measure of the deviation of the distributions from the Maxwellian form. Once the reference state is well characterized, as a second step we obtain exact expressions for the mass flux, the pressure tensor, and the heat flux  in the first order of the hydrodynamic gradients (Navier-Stokes order). From these expressions we identify the seven relevant transport coefficients of the problem, namely the mutual diffusion $D$, Eq.\ (\ref{4.8}), the pressure diffusion $D_p$, Eq.\ (\ref{4.9}), the thermal diffusion $D'$, Eq.\ (\ref{4.10}), the shear viscosity $\eta$, Eqs.\ (\ref{4.19}) and (\ref{4.22}), the Dufour coefficient $D''$, the pressure energy coefficient $L$, and the thermal conductivity $\lambda$ as given by Eqs.\ (\ref{4.34})--(\ref{4.43}), respectively. These expressions are {\em exact} (within the context of IMM) and constitute the main goal of this paper. This contrasts with the previous results derived for IHS, \cite{GD02} where the transport coefficients have been {\em approximately} obtained by considering the leading terms in a Sonine polynomial expansion of the distribution function.

\begin{figure}
\includegraphics[width=0.25 \columnwidth,angle=-90]{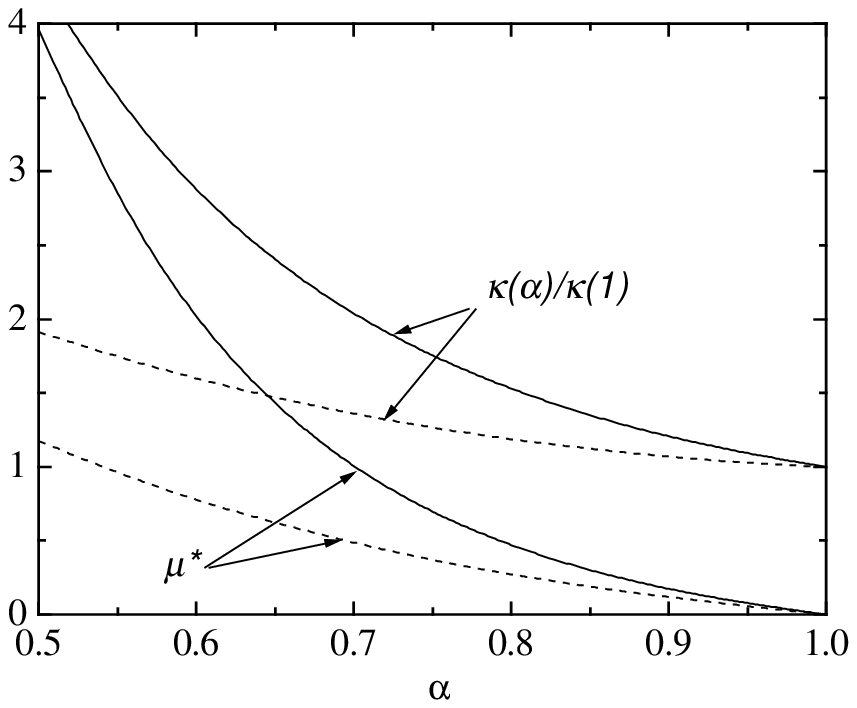}
\caption{Plot of the reduced coefficients $\kappa(\alpha)/\kappa(1)$ and $\mu^*=n\mu/T\kappa(1)$ as a function of the coefficient of restitution $\alpha$ in the three-dimensional case for a single gas.   The solid lines correspond to the exact results obtained for IMM \cite{S03} while the dashed lines are the results derived for IHS in the first Sonine approximation. \cite{BDKS98}
\label{fig9}}
\end{figure}

The purpose of our work is twofold. First, the evaluation of the transport coefficients for mixtures of IMM is worthwhile studying by itself as a simple model to gain some insight into the influence of collisional cooling on the transport properties of the system. Second, the comparison between the exact results for IMM  with the results obtained for IHS by using approximate analytical  methods allows us to assess the degree of reliability of IMM to reproduce the relevant behavior of IHS in the context of granular mixtures. Recent results\cite{G03} derived for multicomponent systems in the simple shear flow problem show a good agreement between both interaction models over a wide range values of the parameter space.

To make contact with the results obtained for IHS\cite{GD02} one needs to fix the collision frequencies $\omega_{rs}$. These quantities can be seen as free parameters of the model to be adjusted to optimize the agreement with IHS. As in the study made in Ref.\ \onlinecite{G03}, here we have chosen $\omega_{rs}$ to reproduce the cooling rates $\zeta_{rs}$ of IHS in the local equilibrium approximation, Eq.\ (\ref{2.31}). An exploration of the dependence of the seven transport coefficients on the full parameter space (mass ratio, diameters, concentrations, coefficients of restitution) is straightforward but perhaps beyond the scope of this presentation. Here, we have focused on the dependence of the coefficients $D$, $D_p$, $D'$, and $\eta$ on the common restitution coefficient $\alpha$ for different values of composition, sizes, and masses. The comparison with known results for IHS\cite{GD99,MG03bis,GM04} has been illustrated in Figs.\ \ref{fig3}--\ref{fig8}. In addition, Monte Carlo simulations have been also included in some plots. The comparison shows that, in general, the IMM predictions are reasonably good for not too large dissipation (say, $\alpha\geq 0.8$), especially for the transport coefficients associated with the mass flux (first-degree velocity moment). The discrepancies between both interaction models increase in the case of the shear viscosity coefficient, which is related to a second-degree moment (pressure tensor). However, the IMM results still capture qualitatively well the dependence of $\eta$ on $\alpha$ for IHS since the discrepancies between both models are about 4\% at $\alpha=0.8$ and about 9\% at $\alpha=0.5$.  As in the monocomponent case, \cite{S03} more significant disagreement between IMM and IHS are expected when one compares higher degree moments, such as the heat flux. To show these discrepancies, in Fig.\ \ref{fig9} we plot the reduced coefficients $\kappa(\alpha)/\kappa(1)$ and $\mu^*=n\mu/T\kappa(1)$ for $d=3$ in the single gas case. Here, $\kappa(\alpha)$ is given by Eq.\ (\ref{4.43.2}) while the coefficient $\mu(\alpha)$ is given by Eq.\ (\ref{4.43.3}). As was emphasized in Ref.\ \onlinecite{S03}, the trends observed for IHS are strongly exaggerated by the IMM, where $\kappa$ and $\mu$ diverge at $\alpha=\frac{1}{9}$.

A simple application of the results obtained in this paper would be the determination of the dispersion relations for the hydrodynamic equations linearized about a homogeneous state. This analysis would allow us to identify the conditions for stability as functions of the wave vector, the dissipation, and the parameters of the mixture. Another possible direction of study would be to check the validity of Onsager's reciprocal relations \cite{GM84} between the different transport coefficients associated with the mass and heat fluxes. Since the system is not time reversal invariant, violation of Onsager's relations is expected for inelastic collisions. The goal would be to assess the influence of dissipation on such violation. We plan to work on these issues in the next future.

\appendix

\section{Collisional moments in the inelastic Maxwell model}
\label{appA}

In this Appendix we will determine the collisional integrals of $m_r{\bf V}$, $m_r{\bf V}{\bf V}$, and $m_rV^2{\bf V}$ appearing in the evaluation of the transport coefficients. The two first integrals were already obtained by one of the authors in a previous paper \cite{G03} on the analysis of rheological properties. Now, for the sake of completeness, we display their explicit expressions:
\begin{equation}
\label{a1}
\int d{\bf v} m_r{\bf V}J_{rs}[f_r,f_s]=-\frac{w _{rs}}{\rho_sd}\mu_{sr}(1+\alpha _{rs})
\left(\rho_s{\bf j}_r-\rho_r{\bf j}_s\right),
\end{equation}
\begin{eqnarray}
\label{a2}
\int d{\bf v} m_r{\bf V}{\bf V}J_{rs}[f_r,f_s]&=& -\frac{w _{rs}}{\rho_sd}\mu_{sr}
(1+\alpha _{rs})\left\{2\rho_s{\sf P}_r-\left(
{\bf j}_r{\bf j}_s+{\bf j}_s{\bf j}_r\right)\right. \nonumber\\
& &-\frac{2}{d+2}\mu_{sr}(1+\alpha _{rs})\left[\rho_s{\sf P}_r+\rho_r{\sf P}_s-
\left({\bf j}_r{\bf j}_s+{\bf j}_s{\bf j}_r\right)\right.\nonumber\\
& &
\left.\left.+\left[\frac{d}{2}\left(\rho_rp_s+\rho_sp_r\right)-{\bf j}_r\cdot {\bf j}_s\right]\openone
\right]\right\},
\end{eqnarray}
where $\openone$ is the $d\times d$ unit tensor. It only remains here to get the collisional integral ${\bf Q}_{rs}$ corresponding to the heat flux:
\begin{equation}
\label{a2bis}
{\bf Q}_{rs}\equiv\int d{\bf v} \case{1}{2}m_rV^2{\bf V}J_{rs}[{\bf v}|f_r,f_s].
\end{equation}
To simplify its calculation, a useful identity for an arbitrary function $h({\bf v})$ is given by
\begin{equation}
\label{a3}
\int d{\bf v}_1 h({\bf v}_1)J_{rs}[{\bf v}_1|f_r,f_s]=\frac{w _{rs}}{n_s\Omega_d}
\int \,d{\bf v}_{1}\,\int \,d{\bf v}_{2}f_{r}({\bf v}_{1})f_{s}({\bf v}_{2})
\int d\widehat{\boldsymbol{\sigma}}\,\left[h({\bf v}_1'')-h({\bf v}_1)\right],
\end{equation}
with
\begin{equation}
\label{a4}
{\bf v}_{1}''={\bf v}_{1}-\mu _{sr}(1+\alpha _{rs})(
\widehat{\boldsymbol{\sigma }}\cdot {\bf g}_{12})\widehat{\boldsymbol{\sigma}},
\end{equation}
and ${\bf g}_{12}={\bf v}_1-{\bf v}_2$.  Now, we particularize to $h({\bf V})=\case{1}{2}m_r{\bf V}V^2$. In this case, using (\ref{a3}) one has
\begin{equation}
\label{a4bis}
{\bf Q}_{rs}=\frac{w _{rs}}{n_s\Omega_d}\frac{m_r}{2}
\int \,d{\bf v}_{1}\,\int \,d{\bf v}_{2}f_{r}({\bf v}_{1})f_{s}({\bf v}_{2})
\int d\widehat{\boldsymbol{\sigma}}\,\left(V_1^{''2}{\bf V}_1''-V_1^2{\bf V}_1\right).
\end{equation}
From the scattering rule (\ref{a4}) it follows that
\begin{eqnarray}
\label{a5}
V_1^{''2}{\bf V}_1''-V_1^2{\bf V}_1&=&
-\mu_{sr}(1+\alpha_{rs})(\widehat{\boldsymbol{\sigma }}\cdot {\bf g}_{12})\left\{\left[V_1^2-
2\mu_{sr}(1+\alpha_{rs})(\widehat{\boldsymbol{\sigma }}\cdot {\bf g}_{12})
(\widehat{\boldsymbol{\sigma }}\cdot {\bf V}_1)\right.\right.\nonumber\\
& & \left.+\mu_{sr}^2
(1+\alpha_{rs})^2(\widehat{\boldsymbol{\sigma }}\cdot {\bf g}_{12})^2\right]\widehat{\boldsymbol{\sigma }}\nonumber\\
& &
\left. +\left[2(\widehat{\boldsymbol{\sigma }}\cdot {\bf V}_1)-
\mu_{sr}(1+\alpha_{rs})(\widehat{\boldsymbol{\sigma }}\cdot {\bf g}_{12})\right]{\bf V}_1\right\}.
\end{eqnarray}
To perform the angular integration, one needs the results
\begin{equation}
\label{a6}
\int d\widehat{\boldsymbol{\sigma}}\,(\widehat{\boldsymbol{\sigma}}\cdot {\bf g}_{12})^k
\widehat{\boldsymbol{\sigma}}=B_{k+1} g_{12}^{k-1}{\bf g}_{12},
\end{equation}
\begin{equation}
\label{a7}
\int d\widehat{\boldsymbol{\sigma}}\,(\widehat{\boldsymbol{\sigma}}\cdot {\bf g}_{12})^k \widehat{\boldsymbol{\sigma}}\widehat{\boldsymbol{\sigma}}
=\frac{B_{k}}{k+d} g_{12}^{k-2}\left(k{\bf g}_{12}{\bf g}_{12}+g_{12}^2
\openone\right),
\end{equation}
where \cite{EB02}
\begin{equation}
\label{a8}
B_k=\int d\widehat{\boldsymbol{\sigma}}\,(\widehat{\boldsymbol{\sigma}}\cdot
{\widehat{\bf g}}_{12})^k=\Omega_d\pi^{-1/2}\frac{\Gamma\left(\frac{d}{2}\right)
\Gamma\left(\frac{k+1}{2}\right)}{\Gamma\left(\frac{k+d}{2}\right)}.
\end{equation}
Taking into account Eqs.\ (\ref{a5}) and (\ref{a6})--(\ref{a8}), the integration over $\widehat{\boldsymbol{\sigma}}$ in Eq.\ (\ref{a4bis}) leads to
\begin{eqnarray}
\label{a9}
{\bf Q}_{rs}&=&-\frac{w _{rs}}{n_sd(d+2)}\frac{m_r}{2}\mu_{sr}(1+\alpha _{rs})
\int \,d{\bf v}_{1}\,\int \,d{\bf v}_{2}f_{r}({\bf v}_{1})f_{s}({\bf v}_{2})\nonumber\\
&  &\times
\left[(d+2)V_1^2{\bf g}_{12}-4\mu_{sr}(1+\alpha _{rs})
\left({\bf g}_{12}\cdot {\bf V}_1\right){\bf g}_{12}
-(d+4)\mu_{sr}(1+\alpha _{rs})g_{12}^2{\bf V}_1\right. \nonumber\\
& & \left.
+3\mu_{sr}^2(1+\alpha _{rs})^2g_{12}^2{\bf g}_{12}+2(d+2)
\left({\bf g}_{12}\cdot {\bf V}_1\right){\bf V}_{1}\right].
\end{eqnarray}
The corresponding integrations over velocity give the relations
\begin{equation}
\label{a10}
\int d{\bf v}_{1}\int \,d{\bf v}_{2}\frac{m_r}{2}V_1^2{\bf g}_{12}f_{r}({\bf v}_{1})f_{s}({\bf v}_{2})=n_s
{\bf q}_r-\frac{d}{2m_s}p_r{\bf j}_s,
\end{equation}
\begin{equation}
\label{a11}
\int d{\bf v}_{1}\int d{\bf v}_{2}\frac{m_r}{2}({\bf g}_{12}\cdot {\bf V}_1){\bf g}_{12}f_{r}({\bf v}_{1})f_{s}({\bf v}_{2})=n_s{\bf q}_r-\frac{1}{2m_s}\left(dp_r{\bf j}_s+{\sf P}_r\cdot {\bf j}_s-
{\sf P}_s \cdot {\bf j}_r\right),
\end{equation}
\begin{equation}
\label{a12}
\int d{\bf v}_{1}\int d{\bf v}_{2}\frac{m_r}{2}g_{12}^2{\bf V}_1f_{r}({\bf v}_{1})f_{s}({\bf v}_{2})=n_s{\bf q}_r+\frac{1}{2m_s}\left(dp_s{\bf j}_r-2{\sf P}_r \cdot {\bf j}_s\right),
\end{equation}
\begin{equation}
\label{a13}
\int d{\bf v}_{1}\int d{\bf v}_{2}\frac{m_r}{2}({\bf g}_{12}\cdot {\bf V}_1){\bf V}_{1}f_{r}({\bf v}_{1})f_{s}({\bf v}_{2})=n_s{\bf q}_r-\frac{1}{2m_s}{\sf P}_r \cdot {\bf j}_s,
\end{equation}
\begin{eqnarray}
\label{a13bis}
\int d{\bf v}_{1}\int d{\bf v}_{2}\frac{m_r}{2}g_{12}^2{\bf g}_{12}& & f_{r}({\bf v}_{1})f_{s}({\bf v}_{2})=n_s{\bf q}_r-\frac{m_r}{m_s}n_r{\bf q}_r\nonumber\\
& &
-\frac{1}{2m_s}\left(dp_r{\bf j}_s-dp_s{\bf j}_r+2{\sf P}_r \cdot {\bf j}_s
-2{\sf P}_s \cdot {\bf j}_r\right).
\end{eqnarray}
From Eqs.\ (\ref{a10})--(\ref{a13}) one finally gets the expression of ${\bf Q}_{rs}$:
\begin{eqnarray}
\label{a14}
{\bf Q}_{rs}&=&\frac{\omega_{rs}}{\rho_s}\frac{\mu_{sr}}{d(d+2)}(1+\alpha_{sr})\left\{
\left[\mu_{sr}(1+\alpha_{rs})\left(d+8-3\mu_{sr}(1+\alpha_{rs})\right)-3(d+2)\right]\rho_s{\bf q}_r\right.\nonumber\\
& &
+3\mu_{sr}^2(1+\alpha_{sr})^2\rho_r{\bf q}_s+\frac{d}{2}\left[
\mu_{sr}(1+\alpha_{rs})\left(3\mu_{sr}(1+\alpha_{rs})-4\right)+d+2\right]p_r
{\bf j}_s\nonumber\\
& &
+\frac{d}{2}\mu_{sr}(1+\alpha_{rs})\left[d+4-3\mu_{sr}(1+\alpha_{rs})\right]p_s
{\bf j}_r\nonumber\\
& &
+\left[\mu_{sr}(1+\alpha_{rs})\left(3\mu_{sr}(1+\alpha_{rs})-(d+6)\right)+d+2\right]
{\sf P}_r\cdot {\bf j}_s\nonumber\\
& &
\left.+\mu_{sr}(1+\alpha_{rs})\left[2-3\mu_{sr}(1+\alpha_{rs})\right]
{\sf P}_s \cdot {\bf j}_r\right\},
\end{eqnarray}
where $p_r=n_rT_r$. In the absence of diffusion and for mechanically equivalent particles, the collisional moment ${\bf Q}_{rs}\equiv {\bf Q}$ reduces to
\begin{equation}
\label{a15}
{\bf Q}=-\omega \frac{(d-1)}{d(d+2)}(1+\alpha)\left[1+\frac{d+8}{d-1}\frac{1-\alpha}{4}\right] {\bf q}.
\end{equation}
This expression coincides with the one previously derived in the monocomponent case. \cite{S03}

Finally, let us evaluate the (dimensionless) collision integral of $v^4$ in the HCS, Eq.\ (\ref{2.28}):
\begin{eqnarray}
\label{a16}
\Lambda_{rs}&\equiv& \int\, d{\bf v^*}v^{*4}J_{rs}^*[\Phi_r,\Phi_s]\nonumber\\
&=&\frac{\omega_{rs}^*}{\Omega_d}\int d{\bf v}_{2}^*
\int d\widehat{\boldsymbol{\sigma }}
\Phi_{r}( v_{1}^*)\Phi_{s}( v_{2}^*)\left(v_1''^{*4}-v_1^{*4}\right),
\end{eqnarray}
where use has been made of the property (\ref{a3}). Henceforth, we will use dimensionless quantities and, for the sake of simplicity, the asterisks will be deleted. The scattering rule (\ref{a4}) gives
\begin{eqnarray}
\label{a17}
v_1''^4-v_1^4&=&2\mu_{sr}^2(1+\alpha_{rs})^2(\widehat{\boldsymbol{\sigma }}\cdot {\bf g}_{12})^2
\left[2(\widehat{\boldsymbol{\sigma }}\cdot {\bf v}_{1})^2+2v_1^2+\frac{\mu_{sr}^2}{2}
(1+\alpha_{rs})^2(\widehat{\boldsymbol{\sigma }}\cdot {\bf g}_{12})^2\right]\nonumber\\
& & -4\mu_{sr}(1+\alpha_{rs})(\widehat{\boldsymbol{\sigma }}\cdot {\bf g}_{12})
(\widehat{\boldsymbol{\sigma }}\cdot {\bf v}_{1})\left[v_1^2+
\mu_{sr}^2(1+\alpha_{rs})^2(\widehat{\boldsymbol{\sigma }}\cdot {\bf g}_{12})^2\right].
\end{eqnarray}
Equations (\ref{a6})--(\ref{a8}) allows the angular integral to be performed with the result
\begin{eqnarray}
\label{a18}
\int d\widehat{\boldsymbol{\sigma }}\left(v_1''^4-v_1^4\right)&=&\frac{4B_2}{d+2}
\mu_{sr}^2(1+\alpha_{rs})^2\left[2({\bf v}_1\cdot {\bf g}_{12})^2+\frac{d+4}{2}g_{12}^2v_1^2+\frac{3}{4}
\mu_{sr}^2(1+\alpha_{rs})^2g_{12}^4\right]\nonumber\\
& & -\frac{4B_2}{d+2}
\mu_{sr}(1+\alpha_{rs})({\bf v}_1\cdot {\bf g}_{12})\left[(d+2)v_1^2+3
\mu_{sr}^2(1+\alpha_{rs})^2g_{12}^2\right].
\end{eqnarray}
Therefore, the collision integral $\Lambda_{rs}$ can be written as
\begin{eqnarray}
\label{a19}
\Lambda_{rs}&=&\frac{\omega_{rs}}{\Omega_d}B_2\mu_{sr}^2(1+\alpha_{rs})^2\int d{\bf v}_{1}\int d{\bf v}_{2}\Phi_r(v_1)\Phi_s(v_2)\nonumber\\
& & \times \left\{\frac{1}{d+2}\left[2(d+8)-12\mu_{sr}(1+\alpha_{rs})+3\mu_{sr}^2(1+\alpha_{rs})^2-\frac{4(d+2)}{\mu_{sr}(1+\alpha_{rs})}\right]v_1^4\right.\nonumber\\
& & \left. +3\frac{\mu_{sr}^2(1+\alpha_{rs})^2}{d+2}v_2^4+\frac{2}{d}\left[d+2-6\mu_{sr}(1+\alpha_{rs})
+3\mu_{sr}^2(1+\alpha_{rs})^2\right]v_1^2v_2^2\right\}.
\end{eqnarray}
Finally, by taking into account that
\begin{equation}
\label{a20}
\int d{\bf v}v^2\Phi_r(v)=\frac{d}{2}\theta_r^{-1},
\end{equation}
we get
\begin{eqnarray}
\label{a20bis}
\Lambda_{rs}&=&
\frac{\mu_{sr}(1+\alpha_{rs})}{d(d+2)}\omega_{rs}^*\left\{3\mu_{sr}^3(1+\alpha_{rs})^3\langle v^{*4}\rangle_{s}\right.\nonumber\\
& & +\left[2d+3\mu_{sr}^2(1+\alpha_{rs})^2-6\mu_{sr}(1+\alpha_{rs})+4\right]\left[\mu_{sr}(1+\alpha_{rs})-2\right] \langle v^{*4}\rangle_{r} \nonumber\\
& & +\frac{d(d+2)}{4}\mu_{sr}\theta_r^{-1}\theta_s^{1}
(1+\alpha_{rs})\left[2d+4+6\mu_{sr}(1+\alpha_{rs})\right.\nonumber\\
& & \left.\left.\times
\left(\mu_{sr}(1+\alpha_{rs})-2\right)\right]\right\}.
\end{eqnarray}
In the one-dimensional case ($d=1$), Eq.\ (\ref{a20bis}) agrees with the results derived in Ref.\ \onlinecite{MP02} for the scalar Maxwell model (i.e., with $\omega_{rs}^* \propto x_s$). In addition, for mechanically equivalent particles, the results of the single gas are recovered, namely, \cite{S03}
\begin{eqnarray}
\label{a21}
\Lambda_{rs}\equiv \Lambda&=&\frac{\omega}{23d(d+2)}(1+\alpha)\left\{\left[12\alpha^2(\alpha-1)+4\alpha(4d+17)-12(3+4d)\right\langle v^4\rangle]\right.\nonumber\\
& & \left. +\frac{d(d+2)}{4}(1+\alpha)(4d-1-6\alpha+3\alpha^2)\right\}.
\end{eqnarray}

\section{Derivation of the transport coefficients}
\label{appB}

In this Appendix we will provide some details on the calculation of the Navier-Stokes transport coefficients appearing in the expressions (\ref{3.1})--(\ref{3.3}) of the irreversible fluxes. Let us consider each flux separately.

\subsection{Mass flux}

To first order, the mass flux ${\bf j}_1^{(1)}$ is defined as
\begin{equation}
\label{4.1}
{\bf j}_1^{(1)}=m_1\int d{\bf v}{\bf V} f_1^{(1)}({\bf v}).
\end{equation}
To get this flux from Eq.\ (\ref{3.14}), we need the collisional integral of $m_1{\bf V}$ which has been evaluated in the Appendix \ref{appA}. From the  linearization of Eq.\ (\ref{a1}), one has the result
\begin{equation}
\label{4.2}
\int  d{\bf v}m_1{\bf V} \left({\cal L}_1f_1^{(1)}+{\cal M}_1f_2^{(1)}\right)=\nu {\bf j}_1^{(1)},
\end{equation}
where $\nu$ is the collision frequency
\begin{equation}
\label{4.3}
\nu=\frac{\rho\omega_{12}}{d\rho_2}\mu_{21}(1+\alpha_{12})=\frac{4}{d}\frac{\rho}{n(m_1+m_2)}\left(\frac{\theta_1+\theta_2}{\theta_1\theta_2}\right)^{1/2}\nu_0(1+\alpha_{12}),
\end{equation}
and use has been made of Eq.\ (\ref{2.32}) in the second equality. Next, we multiply both sides of Eq.\ (\ref{3.14}) by $m_1{\bf V}$ and integrate over ${\bf V}$. The result is
\begin{equation}
\label{4.4}
\left(\partial_t^{(0)}+\nu \right){\bf j}_1^{(1)}=-p\left( \frac{\partial }{\partial x_{1}}
x_1\gamma_1\right) _{p,T}\nabla x_1-\frac{n_1T_1}{p}\left(1-\frac{m_1p}{\rho T_1}\right)\nabla p.
\end{equation}
Note that the temperature ratio $\gamma_1$ depends on the hydrodynamic state through the concentration $x_1$. The functional dependence of $\gamma_1$ on $x_1$ can be obtained from the HCS condition (\ref{2.21}) by using the expressions (\ref{2.18bis}) for the partial cooling rates $\zeta_r^{(0)}$.

The mass flux has the structure given by Eq.\ (\ref{3.1}). Dimensional analysis shows that $D\propto T^{1/2}$, $D_p\propto T^{1/2}/p$, and $D'\propto T^{1/2}$.  Consequently,
\begin{eqnarray}
\label{4.5}
\partial_t^{(0)}{\bf j}_1^{(1)}&=&-\zeta^{(0)}\left(T\partial_T+p\partial_p\right) {\bf j}_1^{(1)}\nonumber\\
&=&\left[\frac{m_1m_2n}{2\rho}\zeta^{(0)}D+\rho(D_p+D') \left( \frac{\partial \zeta^{(0)}}{\partial x_{1}}
\right) _{p,T}\right]\nabla x_1\nonumber\\
& & +\frac{\rho\zeta^{(0)}}{p}\left(\frac{3}{2}D_p+D'\right)\nabla p-\frac{\rho\zeta^{(0)}}{2T}D_p\nabla T.
\end{eqnarray}
Upon deriving this expression use has been made of the identities
\begin{eqnarray}
\label{4.6}
\partial _{t}^{(0)}\nabla T &=&-\nabla \left( T\zeta ^{(0)}\right) =-\zeta
^{(0)}\nabla T-T\nabla \zeta ^{(0)}  \nonumber \\
&=&-\frac{\zeta ^{(0)}}{2}\nabla T-T\left[ \left( \frac{\partial \zeta
^{(0)}}{\partial x_{1}}\right) _{p,T}\nabla x_{1}+\frac{\zeta ^{(0)}}{p}
\nabla p\right] ,
\end{eqnarray}
\begin{eqnarray}
\label{4.7}
\partial _{t}^{(0)}\nabla p &=&-\nabla \left( p\zeta ^{(0)}\right) =-\zeta
^{(0)}\nabla p-p\nabla \zeta ^{(0)}  \nonumber \\
&=&-2\zeta ^{(0)}\nabla p-p\left[ \left( \frac{\partial \zeta ^{(0)}}{
\partial x_{1}}\right) _{p,T}\nabla x_{1}-\frac{\zeta ^{(0)}}{2T}\nabla T
\right] ,
\end{eqnarray}
where we have taken into account that
\begin{eqnarray}
\label{4.7.1}
\nabla\zeta^{(0)}&=&\left(\frac{\partial \zeta^{(0)}}{\partial x_1}\right)_{p,T}\nabla x_1+
\left(\frac{\partial \zeta^{(0)}}{\partial p}\right)_{x_1,T}\nabla p+
\left(\frac{\partial \zeta^{(0)}}{\partial T}\right)_{x_1,p}\nabla T\nonumber\\
&=&\left(\frac{\partial \zeta^{(0)}}{\partial x_1}\right)_{p,T}\nabla x_1+
\frac{\zeta^{(0)}}{p}\nabla p-\frac{\zeta^{(0)}}{2T}\nabla T,
\end{eqnarray}
the two latter terms coming from $\zeta^{(0)}\propto nT^{1/2}=pT^{-1/2}$. Inserting Eq.\ (\ref{4.5}) into Eq.\ (\ref{4.4}), one gets the expressions (\ref{4.8}), (\ref{4.9}), and (\ref{4.10}) for the coefficients $D$, $D_p$, and $D'$,  respectively.

\subsection{Pressure tensor}

The pressure tensor ${\sf P}^{(1)}$ can be written as
\begin{equation}
\label{4.11bis.2}
{\sf P}^{(1)}={\sf P}_1^{(1)}+{\sf P}_2^{(1)},
\end{equation}
where the partial contribution ${\sf P}_r^{(1)}$ to the pressure tensor is
\begin{equation}
\label{4.12}
{\sf P}_r^{(1)}=m_r\int d{\bf v}{\bf V} {\bf V}f_r^{(1)}({\bf v}).
\end{equation}
The linearization of Eq.\ (\ref{a2}) leads to the following expression for the collisional integral of $m_1{\bf V}{\bf V}$:
\begin{equation}
\label{4.13}
\int  d{\bf v}m_1{\bf V} {\bf V}\left({\cal L}_1f_1^{(1)}+{\cal M}_1f_2^{(1)}\right)=\tau_{11} {\sf P}_1^{(1)}+\tau_{12} {\sf P}_2^{(1)},
\end{equation}
where
\begin{equation}
\label{4.14}
\tau_{11}=\frac{\omega_{11}}{d(d+2)}(1+\alpha_{11})(d+1-\alpha_{11})+2\frac{\omega_{12}}{d}\mu_{21}
(1+\alpha_{12})\left[1-\frac{\mu_{21}(1+\alpha_{12})}{d+2}\right],
\end{equation}
\begin{equation}
\label{4.15}
\tau_{12}=-2\frac{\omega_{12}}{d(d+2)}\frac{\rho_1}{\rho_2}\mu_{21}^2
(1+\alpha_{12})^2.
\end{equation}
Now, we multiply both sides of Eq.\ (\ref{3.14}) (with $r=1$) by $m_1{\bf V}{\bf V}$ and integrate over ${\bf V}$ to get
\begin{equation}
\label{4.16}
\left(\partial_t^{(0)}+\tau_{11} \right)P_{1,ij}^{(1)}+\tau_{12} P_{2,ij}^{(1)}=-p_1\Delta_{ijk\ell}\nabla_ku_{\ell},
\end{equation}
where $p_1=n_1T_1$ and
\begin{equation}
\label{4.17}
\Delta_{ijk\ell}\equiv \delta_{ik}\delta_{j\ell}+\delta_{i\ell}\delta_{jk}-\frac{2}{d}\delta_{ij}\delta_{k\ell}.
\end{equation}
A similar equation can be obtained for ${\sf P}_2^{(1)}$ from (\ref{4.16}) by interchanging $1 \leftrightarrow 2$. The solution to Eq.\ (\ref{4.16}) (and its corresponding counterpart) has the form
\begin{equation}
\label{4.18}
P_{r,ij}^{(1)}=-\eta_r \Delta_{ijk\ell}\nabla_ku_{\ell}.
\end{equation}
According to Eq.\ (\ref{3.3}), the shear viscosity coefficient $\eta$ is given in terms of the coefficients $\eta_r$ by
\begin{equation}
\label{4.19.1}
\eta=\eta_1+\eta_2.
\end{equation}
Dimensional analysis requires that $\eta_r\propto T^{1/2}$ and so,
\begin{equation}
\label{4.20}
\partial_t^{(0)}{\sf P}_r^{(1)}=-\frac{\zeta^{(0)}}{2}{\sf P}_r^{(1)}.
\end{equation}
Insertion of this relation into Eq.\ (\ref{4.16}) yields the following set of coupled equations for the two coefficients $\eta_r$:
\begin{equation}
\label{4.21}
\left(
\begin{array}{cc}
\tau_{11}-\frac{1}{2}\zeta^{(0)}& \tau_{12}\\
\tau_{21}&\tau_{22}-\frac{1}{2}\zeta^{(0)}
\end{array}
\right)
\left(
\begin{array}{c}
\eta_{1}\\
\eta_{2}
\end{array}
\right)
=\left(
\begin{array}{c}
p_1\\
p_2
\end{array}
\right).
\end{equation}
Its solution is given by Eq.\ (\ref{4.22}).

\subsection{Heat flux}

The heat flux ${\bf q}^{(1)}$ can be written as
\begin{equation}
\label{4.23}
{\bf q}^{(1)}={\bf q}_1^{(1)}+{\bf q}_2^{(1)},
\end{equation}
where the partial contribution ${\bf q}_r^{(1)}$  is given by
\begin{equation}
\label{4.24}
{\bf q}_r^{(1)}=\frac{m_r}{2}\int d{\bf v}V^2 {\bf V}f_r^{(1)}({\bf v}).
\end{equation}
To get the explicit expressions for the fluxes ${\bf q}_r^{(1)}$ we proceed in a similar way as in the case of the pressure tensor. First,  linearization of Eq.\ (\ref{a14}) leads to
\begin{equation}
\label{4.25}
\int  d{\bf v}\frac{m_1}{2}V^2 {\bf V}\left({\cal L}_1f_1^{(1)}+{\cal M}_1f_2^{(1)}\right)=\beta_{11} {\bf q}_1^{(1)}+\beta_{12} {\bf q}_2^{(1)}+A_{12} {\bf j}_1^{(1)},
\end{equation}
where
\begin{eqnarray}
\label{4.27}
\beta_{11}&=&-\frac{\omega_{11}}{4}\frac{(1+\alpha_{11})}{d(d+2)}\left[\alpha_{11}(d+8)-5d-4\right]-\omega_{12}\mu_{21}\frac{(1+\alpha_{12})}{d(d+2)}\nonumber\\
& & \times \left\{\mu_{21}(1+\alpha_{12})\left[d+8-3\mu_{21}(1+\alpha_{12})\right]-3(d+2)\right\},
\end{eqnarray}
\begin{equation}
\label{4.28}
\beta_{12}=-3\omega_{12}\mu_{21}^3\frac{(1+\alpha_{12})^3}{d(d+2)}\frac{\rho_1}{\rho_2},
\end{equation}
\begin{eqnarray}
\label{4.29}
{A}_{12}&=&-\frac{\omega_{11}}{8}\frac{(1+\alpha_{11})}{d(d+2)}\left[\alpha_{11}(d^2-2d-8)+3d(d+2)\right]\frac{T_1}{m_1}-\frac{\omega_{12}}{2}\mu_{21}\frac{(1+\alpha_{12})}{d}\nonumber\\
& & \times
\left\{ \mu_{21}(1+\alpha_{12})\left[d-3\mu_{21}(1+\alpha_{12})+2\right]\frac{T_2}{m_2}\right.
\nonumber\\
& & \left.
-\frac{x_1}{x_2}\left[d+3\mu_{21}^2(1+\alpha_{12})^2-6\mu_{21}(1+\alpha_{12})+2\right]\frac{T_1}{m_2}\right\}.
\end{eqnarray}
Upon writing Eq.\ (\ref{4.25}) use has been made of the relation ${\bf j}_1^{(1)}=-{\bf j}_2^{(1)}$. The corresponding expressions for $\beta_{22}$, $\beta_{21}$, and $A_{21}$ can be easily obtained from Eqs.\ (\ref{4.27})--(\ref{4.29}) by the change $1 \leftrightarrow 2$.  From Eq.\ (\ref{3.14}), one gets
\begin{eqnarray}
\label{4.31}
\left(\partial_t^{(0)}+\beta_{11} \right){\bf q}_1^{(1)}+\beta_{12}{\bf q}_2^{(1)}&=&-{A}_{12}
{\bf j}_1^{(1)}-\frac{d+2}{2}\frac{nT^2}{m_1}\frac{\partial}{\partial x_1}\left[\left(1+\frac{c_1}{2}\right)x_1\gamma_1^2\right]\nabla x_1\nonumber\\
& & -\frac{d+2}{2}\frac{n_1T_1^2}{m_1p}
\left(1-\frac{m_1p}{\rho T_1}+\frac{c_1}{2}\right)\nabla p\nonumber\\
& & -\frac{d+2}{2}\frac{n_1T_1^2}{m_1T}\left(1+\frac{c_1}{2}\right)\nabla T,
\end{eqnarray}
where
\begin{equation}
\label{4.32}
c_r=\frac{2}{d(d+2)}\frac{m_r^2}{n_rT_r^2}\int d{\bf v} V^4 f_r^{(0)}-2.
\end{equation}
The coefficients $c_r$ have been obtained in Section \ref{sec3}.

The solution to Eq.\ (\ref{4.31}) can be written as
\begin{equation}
\label{4.33}
{\bf q}_r^{(1)}=-T^2D_r''\nabla x_1-L_r\nabla p-\lambda_r\nabla T.
\end{equation}
The total heat flux defines the transport coefficients $D''$, $L$, and $\lambda$ through Eq.\ (\ref{3.2}). According to Eqs.\ (\ref{4.23}) and (\ref{4.33}), these transport coefficients  are given in terms of their partial contributions $D_r''$, $L_r$, and $\lambda_r$ as
\begin{equation}
\label{4.34.1}
D''=D_1''+D_2'', \quad L=L_1+L_2,\quad \lambda=\lambda_1+\lambda_2.
\end{equation}

From dimensional analysis, $D_r''\propto T^{-1/2}$, $L_r\propto T^{3/2}/p$, and $\lambda_r\propto T^{1/2}$. Consequently,
\begin{eqnarray}
\label{4.35}
\partial_t^{(0)}{\bf q}_r^{(1)}&=&\left[\frac{3}{2}\zeta^{(0)}T^2D_r''+ \left( \frac{\partial \zeta ^{(0)}}{
\partial x_{1}}\right) _{p,T}(pL_r+T\lambda_r)\right]\nabla x_1\nonumber\\
& & +\zeta^{(0)}\left(\frac{5}{2}L_r+\frac{T\lambda_r}{p}\right)\nabla p+
\zeta^{(0)}\left(\lambda_r-\frac{pL_r}{2T}\right)\nabla T.
\end{eqnarray}
Substitution of Eq.\ (\ref{4.35}) into Eq.\ (\ref{4.31}) and taking into account the expression (\ref{3.1}) for the mass flux, one arrives at the coupled set of equations (\ref{4.36}) for the partial contributions $D_r''$, $L_r$, and $\lambda_r$.

\acknowledgments

Partial support from the Ministerio de Ciencia y Tecnolog\'{\i}a (Spain) through Grant No. FIS2004-01399 (partially financed by FEDER funds) is acknowledged. A. A. is grateful to the Fundaci\'on Ram\'on Areces (Spain) for a predoctoral fellowship.

%\end{references}

\end{document}